\documentclass[fleqn,usenatbib,usedcolumn]{mnras}

\usepackage[british]{babel}             
\usepackage{newtxtext}
\usepackage[slantedGreek]{newtxmath}

\usepackage[T1]{fontenc}

\DeclareRobustCommand{\VAN}[3]{#2}
\let\VANthebibliography\thebibliography
\def\thebibliography{\DeclareRobustCommand{\VAN}[3]{##3}\VANthebibliography}

\usepackage{graphicx}	
\usepackage{amsmath}	
\usepackage{amssymb}	
\usepackage{multirow,makecell}
\usepackage{tabularx}
\usepackage{pdflscape}

\hypersetup{pdfauthor={Y. C. Perrott},
               pdftitle={A 15.5\,GHz detection of the galaxy cluster minihalo in RXJ1720.1+2638},
               pdfkeywords={radiation mechanisms: non-thermal -- galaxies: clusters: individual: RXJ1720.1+2638 -- galaxies: clusters: intracluster medium -- radio continuum: general},
               bookmarksnumbered=true}

\newcommand{\RXJ}{RXJ1720.1$+$2638}

\title[15.5\,GHz detection of the RXJ1720.1+2638 minihalo]{A 15.5\,GHz detection of the galaxy cluster minihalo in RXJ1720.1+2638}

\author[Y. C. Perrott et al.]{Yvette C. Perrott$^{1}$\thanks{E-mail: yvette.perrott@vuw.ac.nz},
Pedro Carvalho$^{2}$,
Patrick J. Elwood$^{3}$,
Keith J.~B.~Grainge$^{4}$,
David A.~Green$^{3}$,
\newauthor
Kamran Javid$^{5,3}$,
Terry Z.~Jin$^{3}$,
Clare Rumsey$^{3}$,
and Richard D. E. Saunders$^{3,5}$
\\
$^{1}$School of Chemical and Physical Sciences, Victoria University of Wellington, PO Box 600, Wellington 6140, New Zealand\\
$^{2}$Cambridge Machines Deep Learning and Bayesian Systems (CMDLABS) Ltd, 22 Wycombe End, Beaconsfield, Buckinghamshire,\\United Kingdom, HP9 1NB\\
$^{3}$Astrophysics Group, Cavendish Laboratory, JJ Thomson Avenue, Cambridge CB3 0HE, UK\\
$^{4}$Jodrell Bank Centre for Astrophysics, School of Physics and Astronomy, The University of Manchester, Manchester M13 9PL, UK\\
$^{5}$Kavli Institute for Cosmology Cambridge, Madingley Road, Cambridge, CB3 0HA, UK\\
}

\date{Accepted XXX. Received YYY; in original form ZZZ}

\pubyear{2021}

\begin{document}
\label{firstpage}
\pagerange{\pageref{firstpage}--\pageref{lastpage}}
\maketitle

\begin{abstract}
\RXJ\ is a cool-core, `relaxed-appearing' cluster with a minihalo previously detected up to 8.4\,GHz, confined by X-ray-detected cold fronts.  We present observations of the minihalo at 13 -- 18\,GHz with the Arcminute Microkelvin Imager telescope, simultaneously modelling the Sunyaev--Zel'dovich signal of the cluster in conjunction with \emph{Planck} and \emph{Chandra} data in order to disentangle the non-thermal emission of the minihalo.  We show that the previously-reported steepening of the minihalo emission at 8.4\,GHz is not supported by the AMI data and that the spectrum is consistent with a single power-law up to 18\,GHz.  We also show the presence of a larger-scale component of the minihalo extending beyond the cold fronts.  Both of these observations could be explained by the `hadronic' or `secondary' mechanism for the production of relativistic electrons, rather than the currently-favoured `re-acceleration' mechanism and/or multiple episodes of jet activity from the active galactic nucleus in the brightest cluster galaxy.
\end{abstract}




\section{Introduction}\label{S:intro}

The massive cluster of galaxies \RXJ\ at $z=0.160$\footnote{We note that the redshift usually given for this cluster is $z=0.164$ or $z=0.1644$, however this seems to originate from a transcription error. \citealt{1992ApJS...80..257E} give $z=0.164$ citing \citealt{1990EObsC...1.....H}, but \citealt{1990EObsC...1.....H} give $z=0.162$ citing \citealt{1988ApJ...325..610H} who measure $z=0.162 \pm 0.004$ consistent with the more recent redshifts.} ($z=0.16010 \pm 0.00018$, optical, \citealt{2011ApJ...741..122O}; $z=0.15990 \pm 0.00054$, X-ray, \citealt{2011MNRAS.410.1797S}) was first detected as an extended X-ray source by \emph{Einstein} \citep{1988ApJ...325..610H}.  It was subsequently identified in the ROSAT Brightest Cluster Survey \citep{1998MNRAS.301..881E} with a measured temperature of 10.2 keV and luminosity of $16.12 \times 10^{44}$ erg\,s$^{-1}$.  It is generally considered a `regular', `relaxed' or `cool-core' cluster based on its X-ray morphology and cooling time estimates (e.g.\ \citealt{2005MNRAS.359.1481B}, \citealt{2017MNRAS.465..858G}, \citealt{2017ApJ...846...51L}).  It has been well studied as part of the Local Cluster Substructure Survey (LoCuSS, e.g.\ \citealt{2016MNRAS.461.3794O}) and Weighing the Giants (WtG, e.g.\ \citealt{2016MNRAS.463.3582M}) samples and is often used for cosmological constraints (e.g.\ \citealt{2015MNRAS.449..199M}, \citealt{2016A&A...594A..24P}).

Despite its relaxed appearance, \citet{2001ApJ...555..205M} found sharp X-ray surface brightness changes (`cold fronts') within $\approx$\,100\,arcsec of the cluster centre, on opposite sides using \emph{Chandra} observations.  They modelled these as a gas density discontinuity (south-east sector) and gas density slope change (north-west sector) and suggested that these features could be explained as the result of either a late-stage merger or the collapse of two co-located density perturbations of different physical scales.  The first scenario was shown to be favoured by \citet{2010PASJ...62..811O} using weak-lensing mass reconstruction maps to demonstrate the presence of a substructure $\approx$\,3\,arcmin to the north of the main cluster.  The presence of this substructure was confirmed by \citet{2011ApJ...741..122O} using a spectroscopic survey of cluster member galaxies, who proposed that the substructure has travelled from the south, passed the core region on the eastern side (generating the cold fronts) and is now headed to the north.

\RXJ\ hosts a radio `minihalo': a steep-spectrum, diffuse radio source enclosing the cluster's central galaxy (brightest cluster galaxy, BCG).  \citet{2008ApJ...675L...9M} and \citet{2014ApJ...795...73G} (hereafter G14) showed that the minihalo emission forms a spiral structure spatially correlated with and bounded by the X-ray cold fronts, in agreement with predictions for turbulent re-acceleration of electrons from past active galactic nucleus (AGN) activity in a `sloshing' cool core by \citet{2013ApJ...762...78Z}.  G14 detected the minihalo emission over a wide frequency range from 0.3 to 8.4\,GHz, and showed a tentative steepening of the radio spectrum between 4.9 and 8.4\,GHz.  A high-frequency spectral break would favour this re-acceleration model for the production of the minihalo emission, since such a break is naturally produced by the balance between energy gain through re-acceleration and energy losses through synchrotron and inverse Compton radiation; this sets a cutoff in the energy distribution of the electrons creating a break in the synchrotron spectrum.  The frequency of the break is determined by the acceleration efficiency.

Here we present observations of the minihalo at 13 -- 18\,GHz; this is, to our knowledge, the highest frequency detection of a minihalo to date.  The paper is structured as follows.  In Section~\ref{S:observations}, we describe the observations and data reduction.  In Section~\ref{S:analysis} we describe the analysis methods used to disentangle the various emission components present.  In Section~\ref{S:discussion} we discuss the implications of our measurement for the particle acceleration mechanism, and in Section~\ref{S:conclusions} we conclude.

Throughout we use J2000 coordinates and use the convention that radio flux density $S$ depends on frequency $\nu$ as $S\propto \nu^{-\alpha}$ with spectral index $\alpha$.  We use $\upLambda$CDM cosmology with $\Omega_{\mathrm m} = 0.3$, $\Omega_{\upLambda} = 0.7$ and $H_0 = 70$\,km\,s$^{-1}$\,Mpc$^{-1}$.  With this cosmology, at redshift $z=0.160$ 1\,arcsec corresponds to 2.78\,kpc.  We use the `cubehelix' colour scheme defined by \citet{2011BASI...39..289G} for radio astronomical images.

\section{Observations}\label{S:observations}

The Arcminute Microkelvin Imager (AMI, \citealt{2008MNRAS.391.1545Z}) is a radio interferometer located near Cambridge, UK and designed for Sunyaev--Zel'dovich (SZ) effect observations.  It consists of two arrays, the Small Array (SA) and Large Array (LA), designed respectively to image the arcminute-scale SZ effect at $\approx$\,3\,arcmin resolution and confusing radio sources at $\approx$\,30\,arcsec resolution.  Both arrays observe over the same frequency band between 13 -- 18\,GHz with a central frequency of 15.5\,GHz.  \RXJ\ was observed on both arrays as part of a campaign to follow up \emph{Planck}-detected clusters of galaxies (e.g.\ \citealt{2015A&A...580A..95P}, \citealt{2016A&A...594A..27P}), using the upgraded correlator described in \citet{2018MNRAS.475.5677H}.  The cluster was observed as a single pointing on the SA, while the LA observations are a 61-point hexagonal raster in order to cover the SA primary beam out to $\approx$\,10 per cent, with a deeper 19-point hexagonal raster at the centre.  The characteristics of the observations are shown in Table~\ref{tab:obs}.

\begin{table*}
	\centering
	\caption{Characteristics of the AMI observations of \RXJ.  All dates are listed while noise-levels are shown for the combined observation set.  The noise level given for the SA single pointing applies to the centre of the pointing, while the noise levels for the LA are fairly constant over the regions given.}
	\label{tab:obs}
	\begin{tabular}{lcccccc} 
		\hline
		Array & Date & Field of view & Strategy & Noise level & $uv$-range & Largest angular scale \\
                      &      & arcmin        &          & $\upmu$Jy\,beam$^{-1}$ & $\lambda$ & arcmin \\
		\hline
		SA & 2016 Dec 08, 10, 27 & 20 & Single pointing & 93 & 200 -- 1200 & 10 \\[1ex]
		\multirow{2}{*}{LA} & \multirow{2}{*}{2016 Dec 08, 24} & 37 & 61-point & 127 & \multirow{2}{*}{600 -- 6000} & \multirow{2}{*}{3.5} \\
		 & & 8 & 19-point & 65 & & \\
		\hline
	\end{tabular}
\end{table*}

\section{Data reduction}\label{S:reduction}

The data were calibrated and imaged in \textsc{CASA}\footnote{\url{https://casa.nrao.edu/}}.  The full-frequency-resolution data (4096 channels) were flagged for narrow-band radio-frequency interference (RFI) using \textsc{rflag} before being binned to 64 channels for subsequent processing.  Observations of the primary calibrator source 3C\,286 nearby in time to the \RXJ\ observations were used to set the flux scale, using the \citet{2013ApJS..204...19P} flux scale along with a correction for the fact that AMI measures $I+Q$ using the integrated polarization properties of 3C\,286 from \citet{2013ApJS..206...16P}; this is an $\approx$\,4.5 per cent correction.  The primary calibration observations were used to set the amplitude and phase bandpasses.  An additional amplitude correction for atmospheric amplitude variations calculated using the `rain gauge' system (see \citealt{2008MNRAS.391.1545Z} for more detail) was also applied.  The bright, point-like sources J1722+2815 (for the LA) and J1716+2616 (for the SA) were observed throughout the observations in an interleaved manner and were used to correct for atmospheric and/or instrumental phase drift.  In the case of the SA, a secondary amplitude correction was also applied using the interleaved calibrator to correct for some residual amplitude drift.  Visibility weights were estimated using the scatter of the visibilities via a local adaptation of the \textsc{statwt} task.

\begin{figure*}
    \includegraphics[trim={1cm 12.2cm 3.5cm 1.5cm},clip,width=0.5\linewidth]{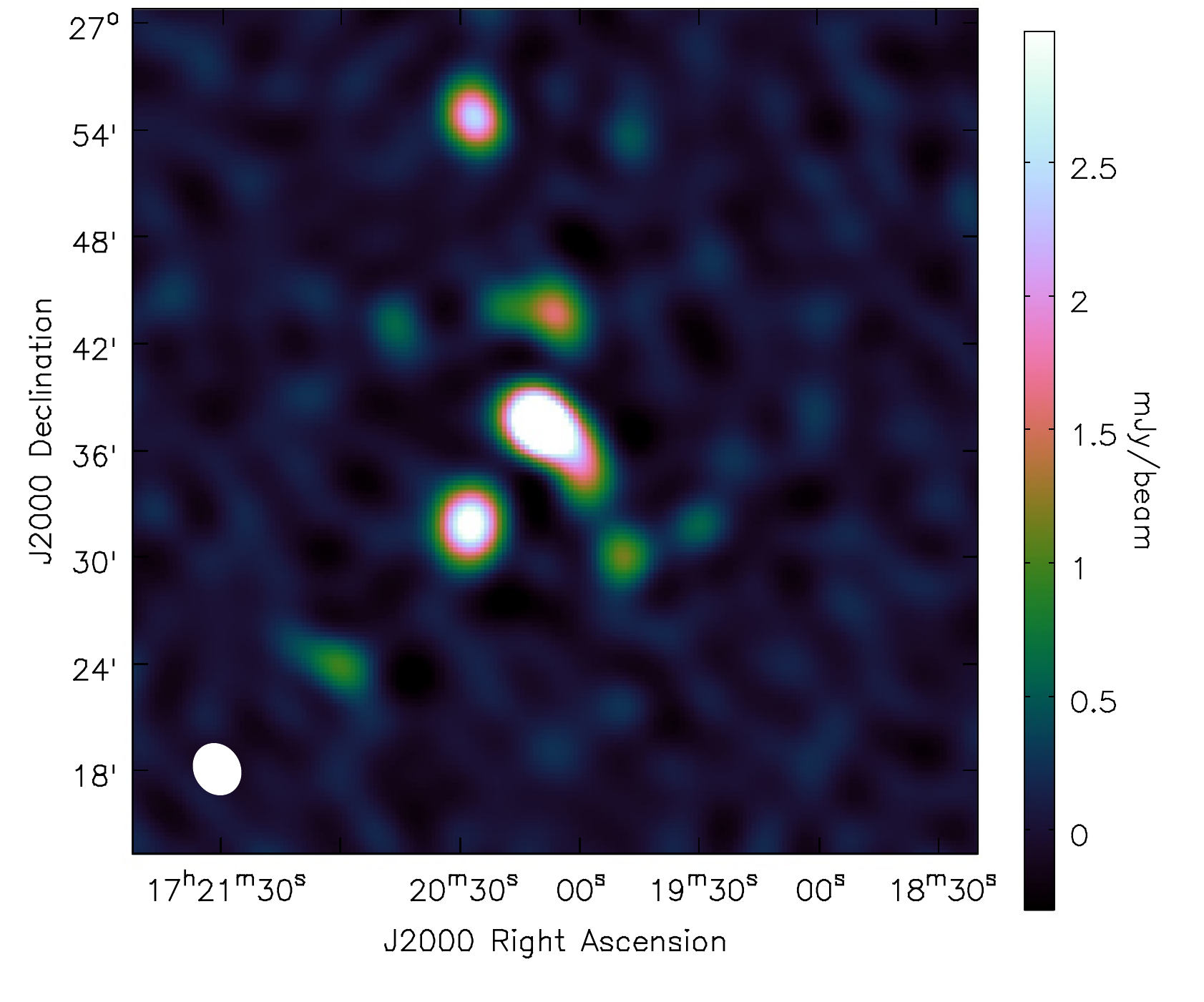}\includegraphics[trim={1cm 12.2cm 3.5cm 1.5cm},clip,width=0.5\linewidth]{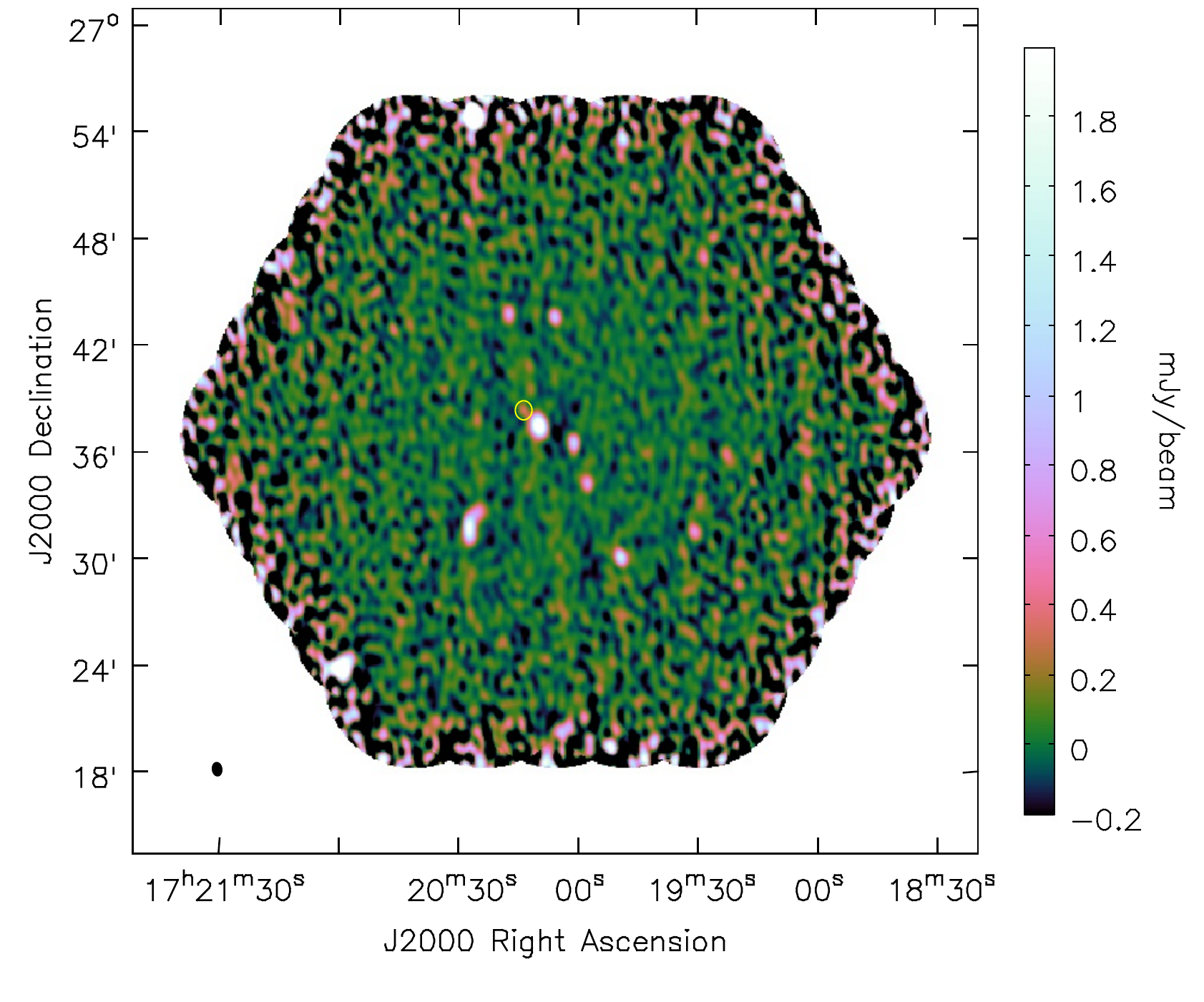}
    \caption{Maps of the \RXJ\ field on the SA (left) and LA (right).  The SA map has not been corrected for the primary beam attenuation.  Both colour scales have been truncated to show low-surface-brightness features; the peak values are 6.5 and 24.7\,mJy\,beam$^{-1}$ respectively for the SA and LA maps.  The synthesised beam is shown as the solid ellipse in the bottom left-hand corner of both maps and has dimensions 178$\times$153\,arcsec$^2$ (SA) and 46$\times$32\,arcsec$^2$ (LA).  The minihalo is the bright source near the map centre.  As discussed in Section~\ref{S:mapplane}, there is a head--tail galaxy immediately to the north-east of the minihalo (its head is barely visible on the LA map and is circled in yellow), and a wide-angle-tail source to the south-east at RA\,$\approx$\,17h20m30s, Dec\,$\approx$\,27d30m which is visibly extended on the LA map.}
    \label{Fi:maps}
\end{figure*}

Mapping was performed using the \textsc{clean} task, using multi-frequency synthesis over the whole bandwidth with \textit{nterms}=2 and natural weighting.  In the case of the LA data, a custom task was used to apply the LA primary beam correction to each pointing centre individually.  A noise map for each pointing centre was created using a sliding box on the residual map; these were also corrected by the primary beam and then converted to weight maps.  The pointing centre weight maps were then used to combine the pointing centre maps into the overall raster map.  An overall raster noise map was also created using the combination of the pointing centre weight maps.  The final maps are shown in Fig.~\ref{Fi:maps}.

\section{Analysis}\label{S:analysis}

\subsection{Map-plane analysis}\label{S:mapplane}

At the centre of the LA map (see Fig.~\ref{Fi:maps}) we can see the extended sources noted by G14.  There is a bright source near the centre coincident with the location of the central part of the minihalo; the head of the head--tail galaxy is barely visible immediately to its north-east (circled in yellow); and the wide-angle-tail source is the visibly extended source to the south-east at RA\,$\approx$\,17h20m30s, Dec\,$\approx$\,27d30m.    A Gaussian fit to the central source shows definite extent with respect to the LA synthesised beam of $\approx$\,46$\times$32\,arcsec$^2$; the deconvolved major-axis FWHM estimate is 30\,arcsec which is consistent with the estimated size from G14.  The 15.5\,GHz emission therefore cannot be attributed simply to the BCG which G14 show has extent $<0.5$\,arcsec.  We do not detect the steeper-spectrum `tail' of the minihalo as seen at lower frequency; from here on when we use the term `minihalo' we refer to the central part only.

We can produce a `raw' estimate of the flux of the central minihalo from the LA raster map.  We use the \textsc{source\_find} software \citep{2011MNRAS.415.2699A} which uses a `flood-fill' technique to estimate an integrated flux density of $4.52 \pm 0.13$ mJy for the central source.  To subtract the BCG contribution, we fit a power-law spectrum to the BCG flux densities given in G14 and obtain a 15.5\,GHz flux density of $0.87 \pm 0.06$\,mJy, implying a minihalo flux density of $3.65 \pm 0.14$\,mJy.  This already implies much less spectral steepening than the 8.44\,GHz measurement from G14 suggests; an extrapolation from the 4.86\,GHz and 8.44\,GHz measurements of G14 predict a 15.5\,GHz flux density of 1.8\,mJy.  However, predictions of the cluster SZ signal using the mass from the second \emph{Planck} SZ catalogue (PSZ2; \citealt{2016A&A...594A..27P}) of $M_{500} = 6.34^{+0.38}_{-0.40}\times 10^{14} M_{\odot}$ show that the shorter baselines of the LA will be contaminated by the (negative) SZ signal and a more detailed analysis is required to isolate the positive minihalo signal.

\subsection{Joint $uv$-plane analysis of SZ and minihalo}

Since the SZ signal contributes significantly only on the very shortest LA baselines, it is difficult to constrain the SZ properties from the LA data.  We therefore incorporate the SA data, which has a better range of angular scales for characterising the arcminute-scale SZ, in the analysis.  We analyse the data in the $uv$-plane rather than the map plane, to properly account for the spatial filtering of the interferometers, after binning the data in $uv$-space to reduce the data volume.  

We begin by considering a simple model consisting of a cluster with a fixed-shape pressure profile, and confusing radio sources represented as point sources.  We then move on to an extended model for the radio minihalo, and allow the shape of the cluster pressure profile to vary to ensure we have accurately disentangled the SZ and minihalo signals.

\subsubsection{AMI-SA analysis}

Our usual strategy to detect SZ signal with AMI is to detect radio sources in the LA data, and model them as point sources in the SA data simultaneously with the cluster model using the Bayesian analysis software \textsc{McAdam}; see, e.g.\ \citet{2009MNRAS.398.2049F}.  This analysis exploits the difference between unresolved radio sources (constant flux density independent of baseline) and the resolved cluster signal (flux density dependent on baseline length) to disentangle the different emission sources.  We perform this analysis as a first pass.  We search for sources in the LA map at $4\sigma$ using \textsc{source\_find} and the noise map described in Section~\ref{S:reduction}, finding a total of 20 sources which are listed in Table~\ref{tab:source_priors}.  To model these in the SA data, in all cases we fix the source position to the LA position.  In cases where the source is detected at $>4\sigma$ in the SA map or is $<5$\,arcmin from the X-ray cluster position, we fit the central flux density and spectral index of the source using a Gaussian prior on the flux density centred on the LA estimate with width of 20 per cent (allowing for possible variability and/or calibration offsets), and the spectral index using an empirical prior based on the 15 -- 22\,GHz spectral index distribution from the 9C survey \citep{2007MNRAS.379.1442W}.  We refer to these sources as `modelled' sources.  Otherwise, we fix the central flux density to the LA value and the spectral index to the median of the spectral index distribution at the appropriate flux density \citep{2013MNRAS.429.2080W}; we refer to these as `fixed' sources.  We use a covariance matrix to account for the remaining source confusion at $<4\sigma_{\mathrm{LA}}$ as well as the primordial CMB anisotropies and (uncorrelated) instrumental noise.

\begin{landscape}
\begin{table}
	\centering
	\caption{Radio sources detected in the AMI maps.  Coordinates are J2000, $(\upDelta) S_{15.5}$ is 15.5\,GHz flux density (error) in mJy, and $\alpha$ is the median spectral index at the flux density of the source from \citet{2013MNRAS.429.2080W}. `Sep' is the distance from the cluster centre, and `type' is an indication of extension in the LA map.  `$>4\sigma_{\mathrm{SA}}$?' indicates whether the source is detected at $>4\sigma$ in the SA data.  Similarly `$>4\sigma_{\mathrm{LA,cen}}$?' indicates whether the source is detected at $>4\sigma$ on the central LA pointing; `NV' indicates that the source is not visible in the central pointing (it is outside the primary beam).  The sources above the horizontal line are detected at $>4\sigma$ on the LA raster map; the bottom two sources are resolved out in the LA data and detected in the SA map only.}
	\label{tab:source_priors}
\begin{tabularx}{\linewidth}{lcccccccccX}
\hline
Source ID & RA & Dec & Sep & Type & $S_{15.5}$ & $\upDelta S_{15.5}$ & $\alpha$ & $>4\sigma_{\mathrm{SA}}$? & $>4\sigma_{\mathrm{LA,cen}}$? & Note \\
 & & & arcmin & & mJy & mJy & & & \\\hline
AMILA J172009+263731 & 17:20:09.98 & +26:37:30.91 & 0.20 & E & 4.40 & 0.13 & 0.79 & \checkmark & \checkmark & BCG + minihalo centre\\
AMILA J172013+263822 & 17:20:13.67 & +26:38:22.31 & 1.23 & P & 0.353 & 0.081 & 0.07 & $\times$ & \checkmark & Head-tail galaxy, external prior\\
AMILA J172001+263632 & 17:20:01.15 & +26:36:32.44 & 2.12 & P & 1.179 & 0.077 & 0.5 & \checkmark & \checkmark & \\
AMILA J172012+264052 & 17:20:12.36 & +26:40:52.36 & 3.31 & P & 0.266 & 0.066 & 0.05 & $\times$ & $\times$ & Dubious detection and leaving a negative residual from fitted subtraction, set to 0\\
AMILA J172025+263751 & 17:20:25.44 & +26:37:51.58 & 3.62 & P & 0.324 & 0.059 & 0.06 & $\times$ & $\times$ & \\
AMILA J171957+263416 & 17:19:57.71 & +26:34:16.61 & 4.23 & P & 0.818 & 0.060 & 0.28 & \checkmark & $\times$ & Tail visible in NVSS/LA, prior widened\\
AMILA J172033+263932 & 17:20:33.72 & +26:39:32.82 & 5.79 & P & 0.277 & 0.069 & 0.05 & $\times$ & NV & Dubious detection and leaving a negative residual from direct subtraction, set to 0\\
AMILA J172005+264337 & 17:20:05.90 & +26:43:37.48 & 6.04 & P & 1.058 & 0.066 & 0.42 & \checkmark & NV & Tail visible in NVSS, prior widened\\
AMILA J172017+264347 & 17:20:17.42 & +26:43:47.76 & 6.43 & P & 0.824 & 0.073 & 0.28 & \checkmark & NV & \\
AMILA J172027+263146 & 17:20:27.23 & +26:31:49.43 & 7.06 & E & 3.87 & 0.46 & 0.79 & \checkmark & NV & WAT\\
AMILA J171949+263007 & 17:19:49.35 & +26:30:07.19 & 8.73 & P & 1.490 & 0.071 & 0.72 & \checkmark & NV & \\
AMILA J172046+264255 & 17:20:46.52 & +26:42:55.75 & 9.87 & P & 0.414 & 0.099 & 0.09 & $\times$ & NV & Large residual from direct subtraction, modelled although $<4\sigma_{\mathrm{SA}}$\\
AMILA J171930+263133 & 17:19:30.80 & +26:31:33.16 & 10.53 & P & 0.701 & 0.091 & 0.22 & $\times$ & NV & \\
AMILA J171922+263552 & 17:19:22.76 & +26:35:52.67 & 10.54 & P & 0.449 & 0.084 & 0.11 & $\times$ & NV & \\
AMILA J171928+264700 & 17:19:28.49 & +26:47:00.07 & 13.07 & P & 0.57 & 0.12 & 0.16 & $\times$ & NV & \\
AMILA J171948+265338 & 17:19:48.78 & +26:53:38.64 & 16.65 & P & 1.71 & 0.21 & 0.79 & $\times$ & NV & \\
AMILA J171857+264356 & 17:18:57.06 & +26:43:56.81 & 17.32 & P & 1.51 & 0.30 & 0.73 & $\times$ & NV & \\
AMILA J172059+262351 & 17:20:59.44 & +26:23:53.60 & 17.74 & E & 8.8 & 1.5 & 0.79 & \checkmark & NV & \\
AMILA J172026+265458 & 17:20:26.44 & +26:54:58.72 & 17.77 & P & 24.87 & 0.73 & 0.67 & $\times$ & NV & \\
AMILA J171951+261926 & 17:19:51.99 & +26:19:26.89 & 18.59 & P & 2.29 & 0.46 & 0.79 & $\times$ & NV & \\
\hline
AMISA J172025+263333 & 17:20:25.06 & +26:33:33.70 & 5.39 & P & 0.815 & 0.095 & - & \checkmark & NV & WAT Northern lobe (SA only)\\
AMISA J172033+263035 & 17:20:33.11 & +26:30:35.49 & 8.83 & P & 0.80 & 0.13 & - & \checkmark & NV & WAT Southern lobe (SA only)\\
\hline
\end{tabularx}

\end{table}
\end{landscape}

Since this cluster is mostly relaxed, a spherically symmetric model consisting of gas in hydrostatic equilibrium with a dark matter halo should describe it well.  Our cluster model consists of a Navarro--Frenk--White (NFW, \citealt{1997ApJ...490..493N}) dark matter halo in hydrostatic equilibrium (HE) with gas with a generalised NFW (GNFW, \citealt{2007ApJ...668....1N}) pressure profile (see \citealt{2012MNRAS.423.1534O} and \citealt{2013MNRAS.430.1344O} for more detail on the model).  The use of this well-motivated physical model, rather than an empirical emission model, allows us to include prior information on the shape of the SZ decrement.  See \citet{KamThesis} for a detailed comparison of physical and empirical cluster models in the context of AMI and \emph{Planck}.  The parameters for the model are the cluster position, redshift, $M_{200}$ (i.e.\ the total mass enclosed by $r_{200}$, where $r_{200}$ is the radius within which the mean cluster density is $200 \times \rho_{\mathrm{crit}}(z)$), $f_{\mathrm{gas,200}}$ (the gas fraction at $r_{200}$; note that $f_{\mathrm{gas}}(r)$ varies throughout the cluster), and the shape parameters $\gamma$, $\alpha$, $\beta$, $c_{500}$ for the GNFW profile.  

The priors on these parameters are listed in Table~\ref{tab:cluster_priors}, where $\mathcal{N}$ and $\mathcal{U}$ refer to normal and uniform distributions respectively.  The AMI observation was centred at the \emph{Planck} position, which has an associated error of $\approx$\,1\,arcmin and is also offset by $\approx$\,1\,arcmin from the BCG.  The X-ray position from the NORAS catalogue \citep{2000ApJS..129..435B} is $<$12\,arcsec from the BCG; given that the cluster is mostly relaxed we would expect the BCG to be close to the intracluster medium (ICM) peak and therefore centre our positional prior at the X-ray position, with a width of 30\,arcsec to allow for a potential SZ--X-ray offset.  The redshift is fixed at $z=0.1644$ also as reported in the NORAS catalogue.  As noted in Section~\ref{S:intro}, this redshift is erroneous but test analyses using the updated redshift of $z=0.160$ show that the difference is too small to significantly impact our SZ analysis.  We initially fix the GNFW shape parameters at the `universal' values defined by \citet{2010A&A...517A..92A}.

\begin{table}
	\centering
	\caption{Priors used on cluster model parameters for the joint cluster--radio source environment $uv$-plane analysis.  $\mathcal{N}$ and $\mathcal{U}$ refer to normal and uniform distributions respectively.  In the case of the GNFW shape parameters, both the initial fixed values and the subsequent uniform priors are shown.}
	\label{tab:cluster_priors}
	\begin{tabular}{lcc}
		\hline
		Parameter & \multicolumn{2}{c}{Prior} \\
		\hline
		RA & \multicolumn{2}{c}{$\mathcal{N}(\mu=260.0386,\sigma=30\,\mathrm{arcsec})$} \\
		Dec & \multicolumn{2}{c}{$\mathcal{N}(\mu=+26.6272,\sigma=30\,\mathrm{arcsec})$} \\
		$z$ & \multicolumn{2}{c}{$\delta(0.1644)$} \\
		$\log_{10}(M_{200})$ & \multicolumn{2}{c}{$\mathcal{U}[14.0,15.7]$} \\
                $f_{\mathrm{gas},200}$ & \multicolumn{2}{c}{$\mathcal{N}(\mu=0.12, \sigma=0.02)$} \\
                $\gamma$ & $\delta(0.3081)$ & $\mathcal{U}[0.0,1.0]$ \\
                $\alpha$ & $\delta(1.0510)$ & $\mathcal{U}[0.1,3.5]$ \\
                $\beta$ & $\delta(5.4905)$ & $\mathcal{U}[3.5,7.5]$ \\
                $c_{500}$ & $\delta(1.177)$ & $\mathcal{U}[0.5,2.0]$ \\
		\hline
	\end{tabular}
\end{table}

\paragraph{Checking the source subtraction}\label{S:source_checks}

We use the fitted radio source properties to subtract the sources from the SA data and visually inspect the resulting subtracted map, to check the quality of the source subtraction.  We find that for two of the `modelled' sources, AMILA J171957+263416 and AMILA J172005+264337, there are significant positive residuals present in the subtracted maps.  Inspection of NVSS maps \citep{1998AJ....115.1693C} of these sources shows possible extent, and the SA posteriors prefer the higher edge of the priors, all of which is an indication that there is flux detected on the SA which is resolved out on the LA.  We therefore widen the priors on the flux densities of these sources to 50 per cent (there is no indication that the sources are resolved on the SA so the point source model is still adequate).  At the location of the `fixed' source AMILA J172046+264255 we also see a significant positive residual although we find no indication of extent on the LA map or in NVSS; we switch this source to `modelled', with a 50 per cent width on flux density.  Noting that AMILA J172013+263822 is the head--tail galaxy, we switch the prior to a tighter one based on a fit to the ancillary data from G14 which gives a 15.5\,GHz flux density prediction of $0.59 \pm 0.09$\,mJy (in agreement with the LA estimate at $2\sigma$) and $\alpha = 0.94 \pm 0.06$.  We add calibration errors in quadrature of 5 per cent and 0.03 to the flux density and spectral index error estimates respectively and use the total errors as our new prior widths.

At the locations of AMILA J172012+264052 ($<4\sigma_{\mathrm{SA}}$ but `modelled' as it is $<5$\,arcmin from the cluster) and AMILA J172033+263932 (`fixed') we see a significant negative residual and note that the posterior for the flux density of AMILA J172012+264052 prefers the lower edge of the prior, approaching 0.  These are marginal detections on the LA map and have no counterparts in lower-frequency maps, so we conclude they are likely to be spurious detections.  We run an extra test analysis, modelling the flux densities of both sources with a uniform prior between $0$ and $S_{\mathrm{LA}} + 3 \times \upDelta S_{\mathrm{LA}}$ and find both posteriors are consistent with 0, supporting the conclusion that these are spurious detections.  We note that AMILA J172012+264052 is relatively close (3.3\,arcmin) to the halo and cluster, and check the posteriors of the source parameters for any correlations with the cluster and halo parameters.  We find none and conclude that, even if real, it is far enough apart to be clearly resolved from them and so it is safe to set the source flux density to 0.  We therefore omit these two sources from here on.

We also note that the lobes of the WAT are not detected on the LA but are detected on the SA subtracted map as $\approx$\,point-like positive residuals.  We run \textsc{source\_find} on the primary-beam-corrected, subtracted SA map to get estimates of their positions and flux densities and add them to the \textsc{McAdam} analysis as extra point sources.  We fix their positions to the map position, put Gaussian priors on flux density using the map estimate and error and use the 9C prior on their spectral index.  These parameters are listed at the bottom of Table~\ref{tab:source_priors}.

Although AMILA J172009+263731, which represents the combination of the BCG and minihalo centre, is subtracted cleanly, its posterior pushes towards the $3\sigma$ upper limit of the LA prior.  Since this source is clearly resolved by the LA and also well-detected on the SA, we change the priors on the central flux and spectral index of this source to uniform priors between 0 and 100\,mJy and $\pm 3$ respectively to avoid bias from missing flux on the LA; i.e.\ we allow the posterior to be purely driven by the SA data.

\paragraph{Cluster detection}\label{S:cluster_detection}

With the adjusted radio source parameter priors described above, we perform two new analysis runs: one with the full cluster + radio point source environment, plus a `null' analysis, where the source properties and priors are identical but the cluster gas mass fraction is fixed to $0$.  Subtracting the sources using the derived parameters from either run shows clean subtraction with no significant residuals in the map.  Comparing the Bayesian evidences of the two runs gives a measure of whether the cluster is detected in the SA data.  In this case $\log(\mathcal{Z}_1/\mathcal{Z}_0)=0.0 \pm 0.4$ (where $\mathcal{Z}_1$ is the Bayesian evidence with one cluster and $\mathcal{Z}_0$ is the Bayesian evidence with no clusters), showing that the AMI-SA data can be explained equally well by a model with or without a cluster.  Given the strong evidence from other wavebands/instruments that there is in fact a massive cluster here, we interpret this to mean that there is not enough information in the SA data to disentangle the cluster signal from the spatially-coincident minihalo signal.  

\begin{figure}
    \includegraphics[width=\linewidth]{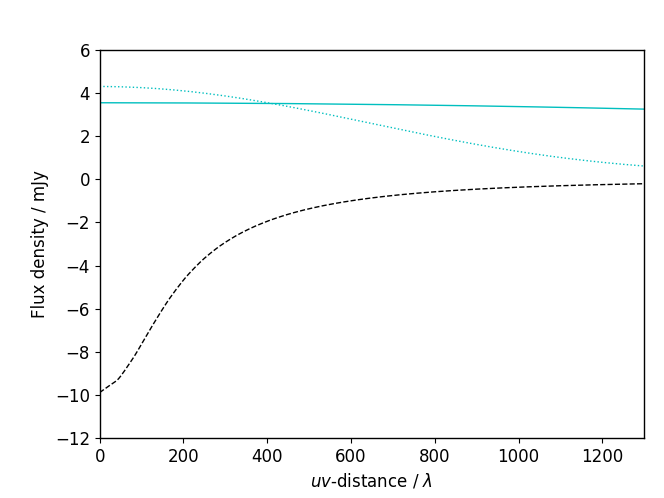}
    \caption{Here we show a representations of the amplitude of three emission components at 15.5\,GHz as a function of baseline length for the SA.  The black dashed line shows a cluster model (as fitted later in the paper), and the solid cyan line shows an exponential halo model with $\approx$\,30\,arcsec extent.  It can be seen that at the resolution of the SA the 30\,arcsec halo is unresolved, the amplitudes showing no dependence on baseline length, while the larger-scale cluster is resolved and therefore has a larger amplitude on the shorter baselines.  This different baseline dependence should allow the two components to be separated in the $uv$-plane analysis.  In contrast, a larger-scale halo, shown with the cyan dotted line, has a similar baseline dependence to the cluster and therefore can approximately cancel out the cluster signal.}
    \label{Fi:SA_uvamps}
\end{figure}

If the minihalo has a $\approx$\,30\,arcsec extent, as measured by G14 and the LA, it should appear point-like to the SA which has angular resolution $\approx$\,3\,arcmin.  In contrast, given the low redshift and high mass of the cluster, the SZ signal will be well-resolved and the two signals should be easily separated.  The fact that the cluster cannot be detected therefore suggests that the minihalo is more extended than expected, meaning both the positive minihalo emission and negative SZ signal have a similar dependence on baseline length.  This makes it far more difficult to disentangle the two emissions using the SA data alone.  This is illustrated in Fig.~\ref{Fi:SA_uvamps}, where the $\approx$\,30\,arcsec component, shown with the solid cyan line, has no significant dependence on baseline length across the baseline lengths observed by the SA.  In contrast, the cluster model, shown with the black dashed line, has a clear dependence on baseline length and therefore should be distinguished from the unresolved structure.  A larger halo model, shown with the dotted cyan line, would have a similar baseline dependence to the cluster and therefore the two signals can approximately cancel out.

\subsubsection{\emph{Planck}-SA analysis}\label{S:Planck-SA}

The additional information required to disentangle the cluster and minihalo emissions can be supplied by \emph{Planck} data.  We use the joint AMI-\emph{Planck} analysis pipeline developed in \citet{2019MNRAS.486.2116P}.  The \emph{Planck} data, being at higher frequencies, will be uncontaminated by the minihalo emission: assuming an extrapolation of the minihalo centre power law spectrum (with no steepening), the total minihalo centre flux density at 100\,GHz will be $\approx$\,1\,mJy with a surface brightness $>150\times$ fainter than the predicted SZ effect signal of the cluster at 100\,GHz.  Even at 217\,GHz, close to the SZ null, the minihalo surface brightness is predicted to be $>15\times$ fainter than the thermal SZ (not accounting for relativistic and kinetic SZ effects which will increase the SZ signal significance).  We therefore do not attempt to account for the minihalo in the \emph{Planck} data and model it as the cluster only.

\begin{figure}
	\includegraphics[width=\linewidth]{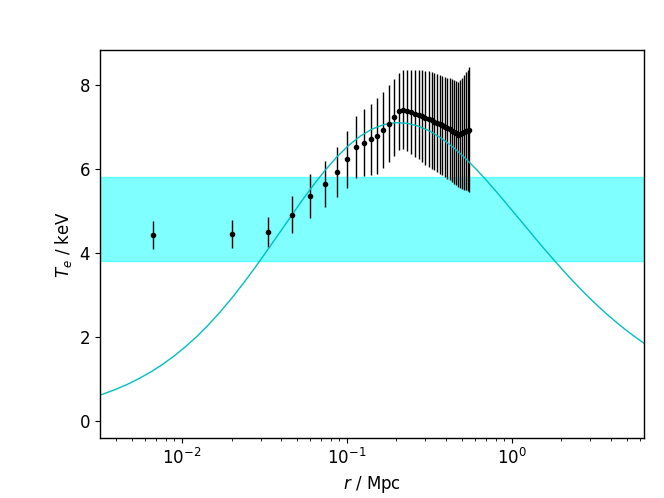}
    \caption{Temperature profile predicted by our hydrostatic equilibrium model, with the best-fit mass from the \emph{Planck}-only analysis (cyan line) compared to the temperature measurements from \emph{Chandra} (black points with errorbars).  The cyan band shows the Compton-$y$-weighted average temperature predicted from the model with the $\pm 1$\,keV prior width.  Although the \emph{Chandra} measurements show an elevated core temperature compared to the model, the prior width allows marginalisation over a physically reasonable range of temperatures.}
    \label{Fi:temperature}
\end{figure}

\begin{figure}
	\includegraphics[width=\linewidth]{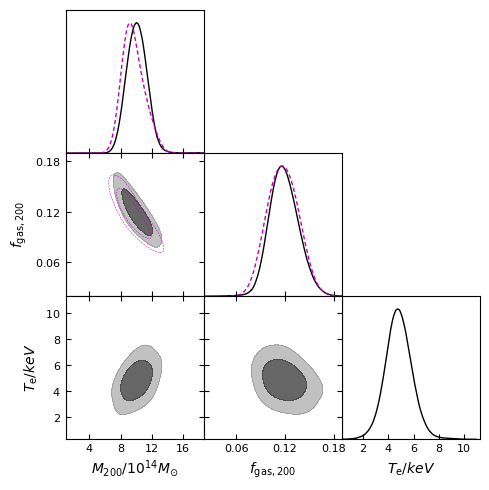}
    \caption{Posterior distributions of the cluster parameters using \emph{Planck}-only analysis, with non-relativistic SZ treatment (unfilled magenta contours, dashed magenta lines) and relativistic SZ treatment (filled grey contours, solid black lines).  Note the non-relativistic SZ treatment is independent of temperature so temperature posteriors are shown for the relativistic case only.  The mean of the mass posterior shifts by $\approx$\,0.4$\sigma$; the posterior on temperature is prior-driven but allows us to marginalize over a physically reasonable range of temperatures.}
    \label{Fi:PL_rSZ}
\end{figure}

We account for relativistic SZ corrections, which can be marginally significant at \emph{Planck} frequencies, as follows.  At the resolution and sensitivity of \emph{Planck} the isothermal approximation is sufficient to account for departures from the non-relativistic SZ spectrum (Perrott et al., in prep.).  Our physical model predicts temperature as a function of radius, so we can use it to calculate an `SZ-average' cluster temperature by calculating the Compton-$y$-weighted spherical average of the temperature, i.e.\ $\int T \, y \, {\rm d}V \, / \, \int y \, {\rm d}V$.  We use this calculated temperature, which is a function of the cluster parameters, as the centre of a Gaussian prior with width of 1\,keV to allow for deviations from the model due to the elevated temperature of the centre compared to the model (see Fig.~\ref{Fi:mazzotta_model}).  We then use relativistic SZ signal calculations from \textsc{SZpack} (\citealt{2012MNRAS.426..510C}, \citealt{2013MNRAS.430.3054C}), averaged over the \emph{Planck} bandpasses provided in the \emph{Planck} Legacy Archive\footnote{\url{https://pla.esac.esa.int}} to create a look-up table of conversion factors between Compton-$y$ and brightness as a function of temperature for each \emph{Planck} HFI frequency channel.  In Fig.~\ref{Fi:PL_rSZ} we show the difference between the posteriors for \emph{Planck}-only analysis with and without the relativistic correction; the mass is very slightly higher in the relativistic case (by $\approx$\,0.4$\sigma$).  The posterior constraint on the temperature is $T_{\rm e} = 4.8 \pm 1.0$\,keV.  The fact that the error bar is the same as the prior width means the temperature is not being constrained by the data, but in comparison to the \emph{Chandra} temperature measurements shown in Fig.~\ref{Fi:temperature} we are confident we have marginalised over a physically reasonable range of temperatures.  We also implement the relativistic SZ calculation for the AMI data, although at AMI frequencies the relativistic correction is not significant ($<2$ per cent at $T_{\rm e} = 5$\,keV).

\begin{figure*}
	\includegraphics[width=\linewidth]{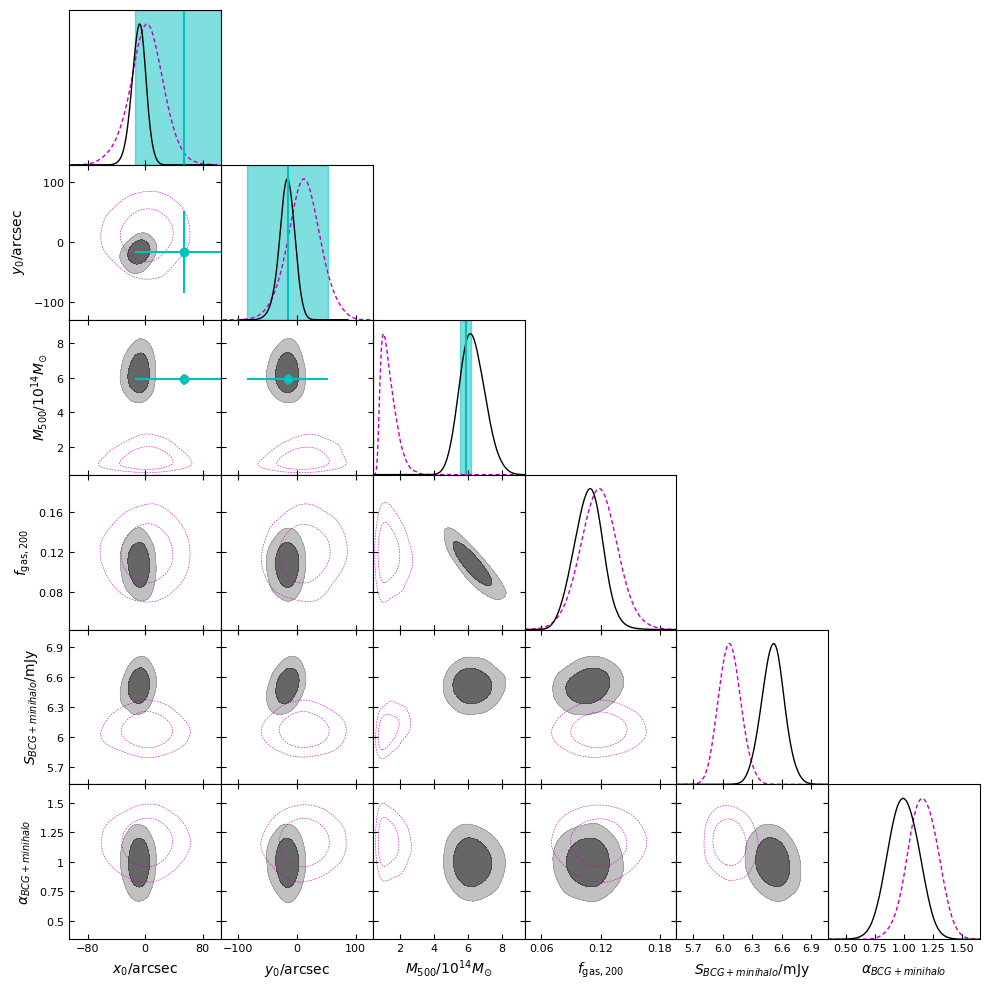}
    \caption{Posterior distributions of the cluster and minihalo parameters, treating the latter as a point source, for AMI-SA only analysis (unfilled magenta contours, dashed magenta lines) and AMI-SA$+$\emph{Planck} analysis (filled grey contours, solid black lines).  $x_{0}$, $y_{0}$ are offsets with respect to the X-ray position.  We show $M_{500}$ rather than our sampling parameter $M_{200}$ for ease of comparison with PSZ2; the estimates from PSZ2 are shown in cyan with error bars (68 per cent confidence level) and vertical lines and bands.  The positional error bars are the 95 per cent confidence limits from the PSZ2 catalogue divided by two.}
    \label{Fi:SA_PL_tri}
\end{figure*}

In the AMI-SA data we model the full radio source $+$ cluster signal as before, keeping the point-source approximation for the minihalo.  In the joint AMI-\emph{Planck} analysis the evidence ratio with respect to the `null' run now clearly favours the presence of a cluster with $\log(\mathcal{Z}_1/\mathcal{Z}_0)=50.6\pm0.4$ (a log-evidence ratio $>2$ is considered `decisive' by \citealt{jeffreys}).  Fig.~\ref{Fi:SA_PL_tri} shows the posterior distributions for the cluster and minihalo parameters in the SA-only and SA$+$\emph{Planck} analysis.  The minihalo flux density is higher in the SA$+$\emph{Planck} analysis showing that in the SA-only analysis the source was filling in the cluster decrement.  We note that there is no degeneracy visible between the cluster mass and the point-source flux density, indicating that the cluster mass estimate is driven by the \emph{Planck} data.  The cluster positional posterior is improved by adding the AMI data however, as shown by comparison with the PSZ2 error bars which are also overlaid on the plot.  The mass estimate is consistent with the PSZ2 estimate even though these were produced using quite different methods (scaling relationships in PSZ2 and hydrostatic mass estimate in this work).  We show best-fit cluster parameters in Table~\ref{tab:cluster_fits}.

\begin{table*}
	\centering
	\caption{Fitted cluster parameters resulting from the combination of various datasets with our fiducial result highlighted in {\bf bold}.  We exclude $f_{g,200}$ which is not constrained by the data and always returns its prior value, $f_{g,200}=0.12\pm0.02$.  `PPv' indicates the GNFW pressure profile parameters are varied.  Errors in RA and Dec are in seconds and arcsec respectively.  Parameters marked with $^{*}$ are fixed.  We note that there is a large degeneracy between $\beta$ and $c_{500}$ as shown in Fig.~\ref{Fi:Chandra_SA_LA_PL}.}
	\label{tab:cluster_fits}
        \renewcommand{\arraystretch}{1.2}
        \footnotesize
\begin{tabularx}{\linewidth}{X>{$}c<{$}>{$}c<{$}>{$}c<{$}>{$}c<{$}>{$}c<{$}>{$}c<{$}>{$}c<{$}}
\hline
 & \mathrm{RA} & \delta & M_{200} & \gamma & \alpha & \beta & c_{500} \\
 & & & 10^{14}\: M_{\odot} & & & & \\\hline
\emph{Planck}-only & 17\:20\:09.3 \pm 1.5 & 26\:36\:45 \pm 22 & 10.1\pm 1.3 & 0.3081^{*} & 1.0510^{*} & 5.4905^{*} & 1.177^{*}\\
SA+\emph{Planck} & 17\:20\:10.25 \pm 0.74 & 26\:37\:15 \pm 13 & 9.2^{+1.0}_{-1.2} & 0.3081^{*} & 1.0510^{*} & 5.4905^{*} & 1.177^{*}\\
SA+LA+\emph{Planck} & 17\:20\:10.8 \pm 1.0 & 26\:37\:18 \pm 16 & 9.9^{+1.1}_{-1.4} & 0.3081^{*} & 1.0510^{*} & 5.4905^{*} & 1.177^{*}\\
SA+LA+\emph{Planck}, PPv & 17\:20\:10.3 \pm 1.3 & 26\:37\:12 \pm 20 & 9.7^{+1.2}_{-1.6} & < 0.425 & 2.01^{+0.68}_{-0.83} & 5.13^{+0.54}_{-1.6} & 1.28\pm 0.34\\
\bf \emph{Chandra}+SA+LA+\emph{Planck}, PPv & \mathbf{17\:20\:10.80 \pm 0.91} & \mathbf{26\:37\:17 \pm 15} & \mathbf{9.48^{+0.90}_{-1.2}} & \mathbf{0.233\pm 0.096} & \mathbf{0.679^{+0.061}_{-0.087}} & \mathbf{5.12^{+0.42}_{-0.70}} & \mathbf{1.28\pm 0.35}\\
\hline
\end{tabularx}

\end{table*}

\subsubsection{Resolved minihalo model}\label{S:resolved_minihalo}

As mentioned in Section~\ref{S:cluster_detection}, the fact that the SA cannot distinguish between the minihalo and cluster emission suggests that the minihalo may be extended on similar scales to the cluster so a more accurate model for the data should include a resolved minihalo.  Unlike for cluster emission, an appropriate physically-motivated model has not been defined for minihalo emission, and such a model derived at lower frequency may not be appropriate for this higher-frequency halo in any case.  \citet{2009A&A...499..679M} showed that an exponential model was a good fit to the surface brightness emission of three separate minihalos, so we also adopt this empirically-motivated exponential model, ie $I(r) = I_{0} \exp(-r/r_{e})$, where $r$ is the projected radius on the sky and $r_{e}$ is the $e$-folding radius.  We cut off the model at $3 \times r_{e}$ and fix the scaling factor $I_{0}$ such that $\int_{0}^{3 r_{e}} I(r) 2\pi r dr = S_{\mathrm{int}}$.  We allow the halo to be ellipsoidal, with minor-to-major axis ratio $f$ and angle of the major axis (defined east of north) $\theta$, i.e.

\begin{multline}
    r^2(x,y) = \left(f \cos^2(\theta)+\sin^2(\theta)/f \right) x^2 \\
+ \left(f \sin^2(\theta)+\cos^2(\theta)/f \right) y^2 + 2 \cos(\theta) \sin(\theta) \left(f-1/f \right) xy
	\label{eq:ellipsoidal}
\end{multline}
where $x$, $y$ are offsets from the minihalo centre.  We assume a single spectral index for the entire minihalo and allow its centre to shift, with a prior centred on the LA map position.  All other parameters have flat priors (see Table~\ref{tab:halo_priors}).  The cluster and radio point source environment models and priors are as before, with the exception that the point source AMILA J172009+263731 which previously represented the BCG + minihalo, now represents the BCG alone with the position shifted to the more accurate FIRST position and an external prior on the flux density and spectral index imposed based on fitting to the spectrum from G14.  To decrease the computational time required, we also fix the flux densities and spectral indices of sources at $>5.5$\,arcmin from the BCG to their previously fitted values.  Since there are no degeneracies between the posteriors for these source parameters and the minihalo/cluster parameters, this does not affect the results.

\begin{table}
	\centering
	\caption{Priors used on minihalo model parameters for the joint cluster--minihalo--radio source environment $uv$-plane analysis.  RA and Dec are the minihalo centre coordinates, $S_{\mathrm{int}}$ is the integrated flux density, $\alpha$ the spectral index, $r_{e}$ the $e$-folding radius, $f$ the minor-to-major axis ratio, and $\theta$ the angle of the major axis, defined east of north.}
	\label{tab:halo_priors}
	\begin{tabular}{lc}
		\hline
		Parameter & Prior \\
		\hline
		RA & $\mathcal{N}(\mu=260.0416,\sigma=10\,\mathrm{arcsec})$ \\
		Dec & $\mathcal{N}(\mu=+26.6253,\sigma=10\,\mathrm{arcsec})$ \\
                $S_{\mathrm{int}}$ & $\mathcal{U}[1, 10]\, \mathrm{mJy}$\\
                $\alpha$ & $\mathcal{U}[-3, 3]$\\
                $r_{e}$ & $\mathcal{U}[5, 50]\, \mathrm{arcsec}$\\
                $f$ & $\mathcal{U}[0.1, 1]$\\
                $\theta$ & $\mathcal{U}[-90, 90]\, \mathrm{deg}$\\
		\hline
	\end{tabular}
\end{table}

\begin{figure*}
	\includegraphics[width=0.5\linewidth]{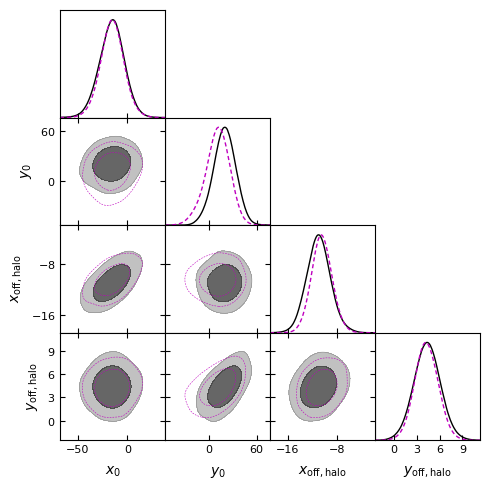}\includegraphics[width=0.5\linewidth]{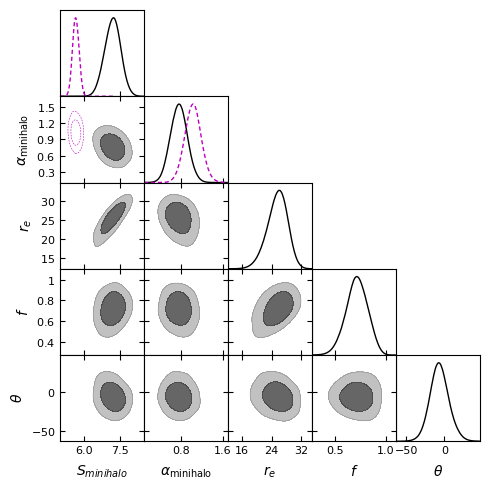}
    \caption{Posterior distributions of the cluster and minihalo parameters, comparing results when treating the latter as a resolved source using an exponential surface brightness model (filled grey contours, solid black lines) and a point source (unfilled magenta contours, dashed magenta lines) with AMI-SA$+$\emph{Planck} analysis.  In the left-hand plot, $x_{0}$, $y_{0}$ are offsets of the cluster centre with respect to the X-ray position and $x_{\mathrm{off,halo}}$, $y_{\mathrm{off,halo}}$ are offsets of the minihalo centre with respect to the X-ray position, all in arcsec.  In the right-hand plot, $S_{\mathrm{minihalo}}$ and $\alpha_{\mathrm{minihalo}}$ are the integrated flux density (in mJy) and spectral index, and $r_{e}$, $f$, $\theta$ are the $e$-folding radius (in arcsec), minor-to-major axis ratio and angle of the major axis east of north (in degrees) for the resolved model.}
    \label{Fi:res_ps_tri}
\end{figure*}

Comparing Bayesian evidences between this model and a point-source model for the halo (with the same radio source environment priors, including BCG, and the same priors on position, flux density and spectral index as the resolved minihalo for a fair comparison), gives a strong preference for the resolved halo model with $\log(\mathcal{Z}_{\mathrm{resolved}}/\mathcal{Z_{\mathrm{point-like}}}) = 8.2 \pm 0.3$.  The Bayesian evidence comparison automatically implements the `Ockham's razor' principle by penalizing the addition of extra parameters that are not warranted by the uncertainties in the data, so this clearly supports an extended minihalo given that the resolved halo model is fitting 3 more parameters ($r_{e}$, $f$, and $\theta$).  We report the fitted halo parameters in Table~\ref{tab:halo_fits} and show posteriors for some of the parameters in Fig.~\ref{Fi:res_ps_tri}.  The model parameters are all well constrained.  A small amount of degeneracy is seen between the cluster position and minihalo position, and the flux density of the minihalo and its $e$-folding radius, but otherwise all parameter constraints are approximately independent.  In particular, there are no significant degeneracies between the minihalo flux density and the cluster parameters showing that the combination of the \emph{Planck} and AMI data is successfully separating the two signals.

\begin{table*}
	\centering
	\caption{Minihalo parameters derived from the combination of various datasets using the exponential model as defined in Section~\ref{S:resolved_minihalo}.  $S_{\mathrm{minihalo}}$ and $\alpha_{\mathrm{minihalo}}$ are the integrated flux density (in mJy) and spectral index, and $r_{e}$, $f$, $\theta$ are the $e$-folding radius (in arcsec), minor-to-major axis ratio and angle of the major axis east of north (in degrees).  `Double' indicates the double exponential minihalo model.  `PPv' indicates that the pressure profile parameters are allowed to vary.  Note that in the case of the VLA 8.44\,GHz fits, flux density and $r_e$ of the larger halo are as constrained by AMI.  Our fiducial result is highlighted in {\bf bold}.}
	\label{tab:halo_fits}
        \scriptsize
        \renewcommand{\arraystretch}{1.2}
\begin{tabular}{l>{$}c<{$}>{$}c<{$}>{$}c<{$}>{$}c<{$}>{$}c<{$}>{$}c<{$}>{$}c<{$}}
\hline
 & \mathrm{RA} & \delta & S_{\mathrm{minihalo}} & \alpha_{\mathrm{minihalo}} & r_{e} & f & \theta \\
 & & & \mathrm{mJy} & & \mathrm{arcsec} & & \mathrm{deg} \\\hline
SA+\emph{Planck} & 17\:20\:10.08 \pm 0.14 & 26\:37\:33.6 \pm 1.8 & 7.19\pm 0.33 & 0.76\pm 0.16 & 26^{+3}_{-2} & 0.72\pm 0.11 & -7\pm 13\\[1ex]
LA+\emph{Planck} & 17\:20\:10.022 \pm 0.038 & 26\:37\:31.94 \pm 0.65 & 4.41\pm 0.21 & 0.81\pm 0.28 & 10.68\pm 0.59 & 0.848\pm 0.072 & -37^{+13}_{-21}\\[1ex]
SA+LA+\emph{Planck} & 17\:20\:10.068 \pm 0.041 & 26\:37\:32.52 \pm 0.61 & 5.52\pm 0.16 & 0.87\pm 0.15 & 13.23\pm 0.48 & 0.813^{+0.055}_{-0.062} & -38\pm 11\\[1ex]
\multirowcell{2}[0ex][l]{SA+LA+\emph{Planck} \\ \quad double} & 17\:20\:10.008 \pm 0.046 & 26\:37\:32.05 \pm 0.76 & 3.72^{+0.37}_{-0.43} & 0.60\pm 0.32 & 9.30\pm 0.99 & 0.842^{+0.10}_{-0.077} & -31^{+17}_{-31}\\
 & 17\:20\:10.27^{+0.43}_{-0.36} & 26\:37\:34.0^{+5.4}_{-4.7} & 4.16\pm 0.47 & 1.15\pm 0.64 & 47.8^{+4.8}_{-10} & 0.71\pm 0.10 & -19^{+11}_{-13}\\[1ex]
\multirowcell{2}[0ex][l]{SA+LA+\emph{Planck} \\ \quad double, PPv} & 17\:20\:10.015 \pm 0.046 & 26\:37\:32.12 \pm 0.76 & 3.84\pm 0.37 & 0.63\pm 0.29 & 9.58^{+0.99}_{-0.87} & 0.859^{+0.11}_{-0.060} & -30^{+18}_{-35}\\
 & 17\:20\:10.13^{+0.48}_{-0.41} & 26\:37\:30.4^{+6.1}_{-5.4} & 3.52\pm 0.62 & 1.66^{+0.82}_{-0.68} & 50^{+8}_{-10} & 0.63\pm 0.12 & -22.8^{+9.9}_{-12}\\[1ex]
\multirowcell{2}[0ex][l]{\bf \emph{Chandra}+SA+LA\\ \bf \quad +\emph{Planck} double, PPv} & \mathbf{17\:20\:10.003 \pm 0.048} & \mathbf{26\:37\:32.05 \pm 0.79} & \mathbf{3.59^{+0.34}_{-0.39}} & \mathbf{0.58\pm 0.35} & \mathbf{9.03\pm 0.97} & \mathbf{> 0.811} & \mathbf{-29^{+17}_{-40}}\\
 & \mathbf{17\:20\:10.32 \pm 0.38} & \mathbf{26\:37\:36.1 \pm 5.4} & \mathbf{4.38^{+0.46}_{-0.41}} & \mathbf{0.92\pm 0.60} & \mathbf{43.4^{+3.6}_{-6.2}} & \mathbf{0.754\pm 0.099} & \mathbf{-18^{+13}_{-15}}\\[1ex]
VLA 8.44\,GHz & 17\:20\:09.974 \pm 0.062 & 26\:37\:31.94 \pm 0.97 & 11.3\pm 1.0 &  & 11.50\pm 0.90 & > 0.874 & < 6.38\\[1ex]
\multirowcell{2}[0ex][l]{VLA 8.44\,GHz double} & 17\:20\:09.960 \pm 0.058 & 26\:37\:31.91 \pm 0.79 & 10.32\pm 0.93 &  & 11.00^{+0.76}_{-0.87} & 0.880^{+0.094}_{-0.052} & < -14.3\\
 & 17\:20\:10.32 \pm 0.65 & 26\:37\:29.6^{+9.7}_{-9.0} & 7.3^{+1.8}_{-3.1} & 0.73\pm 0.58 & 44\pm 5 & --- & 10^{+47}_{-41}\\[1ex]
\hline
\end{tabular}

\end{table*}

\subsubsection{\emph{Planck}-SA-LA analysis}

If the exponential surface brightness model is a good model for the minihalo, the difference between the measured SA flux density (7.1\,mJy in the \emph{Planck}$+$SA analysis) and LA flux density (3.6\,mJy measured from the map) should be reconciled in the $uv$-plane analysis.  We therefore now consider a similar analysis but using the \emph{Planck}$+$LA data.  We use only the central pointing of the LA raster, since it contains the most information on the cluster and halo.  The noise level on this pointing is $\approx$\,63\,$\upmu$Jy beam$^{-1}$.  Only six of the radio sources are visible inside the primary beam of this pointing; we model the flux density and spectral index of three of them (detected at $>4\sigma$ in the pointing) simultaneously with the cluster and minihalo parameters and fix the others to the values derived from the overall LA raster map.  These sources are indicated in Table~\ref{tab:source_priors}.  We use the same priors on the cluster and minihalo parameters as for the \emph{Planck}-SA analysis (Tables~\ref{tab:cluster_priors} and \ref{tab:halo_priors}).  The results are reported in Table~\ref{tab:halo_fits}; the fitted flux density is higher than the original map flux density but is lower than the SA result and inconsistent at $\approx$\,7$\sigma$, and the $e$-folding radius is much smaller ($\approx$\,10\,arcsec compared to $\approx$\,25\,arcsec) and incompatible at $\approx$\,5$\sigma$.  This inconsistency indicates that the exponential surface brightness model is not a good model for the emission over the total range of angular scales probed by the LA and SA.

We finally put the three datasets together and jointly fit the cluster, minihalo and radio source parameters.  The discrepancies between SA and LA radio source flux densities due to source extent discussed in Section~\ref{S:source_checks} do not impact the analysis since all of the resolved sources are outside of the field of view of the LA central pointing except for AMILA J171957+263416, which at 4.3\,arcmin separation from the BCG does not affect the halo or cluster fitting results.  The joint result is reported in  Table~\ref{tab:halo_fits} and as expected the flux density and $r_{e}$ lie between the individual \emph{Planck}$+$SA and \emph{Planck}$+$LA results, pushed toward the LA values due to the lower noise in the LA data.  Subtracting the jointly fitted model from the SA and LA data leaves clear residuals at the minihalo position in both maps, confirming that this model cannot account for both sets of observations.  

To improve the fit, we test a model consisting of two exponential-surface-brightness haloes, both centred at the same position.  We use the same priors on all parameters but to avoid mode-swapping (i.e.\ to keep the halo labelled `1' uniquely associated with the smaller-angular-scale halo, and vice-versa) we set the prior on $r_{e,1}$ as $\mathcal{U}[5, 15]$\,arcsec (i.e.\ to encompass the \emph{Planck}$+$LA posterior) and on $r_{e,2}$ as $\mathcal{U}[15,100]$\,arcsec (i.e.\ to encompass the \emph{Planck}$+$SA posterior and allow some extra width as the posterior was being compressed by the initial 50\,arcsec boundary in the joint analysis).  There is a clear preference for this model when all three datasets are analysed together, with $\log(\mathcal{Z}_{\mathrm{double}}/\mathcal{Z_{\mathrm{single}}}) = 24.8 \pm 0.4$.  Subtracting this model we see no evidence for residual features in either the SA or LA residual maps, aside from a small undersubtraction of AMILA J171957+263416 in the SA map.  The results for the joint fit using the double-halo model are reported in Table~\ref{tab:halo_fits}.

\subsubsection{Pressure profile variation}

We now consider the possibility that the pressure profile of the cluster does not conform to the `universal' pressure profile (UPP) shape.  \citet{2010A&A...517A..92A} showed that their morphologically disturbed and cool-core cluster subsamples had slightly different pressure profile shapes, with the `universal' shape being the average, so we might expect that \RXJ\ conforms more closely to the cool-core shape.  Since the cluster SZ decrement and minihalo exist on the same scales, an inaccurate cluster shape could bias the minihalo flux density.  To test this we rerun the joint \emph{Planck}-SA-LA analysis, with the double-halo model, now allowing the GNFW shape parameters to vary with the uniform priors shown in the third column of Table~\ref{tab:cluster_priors}\footnote{We note that we do not actually vary $\gamma$ in the \emph{Planck} likelihood calculation, since it does not affect the low-resolution \emph{Planck} model significantly; we verify this by comparing \emph{Planck}-only analysis with $\alpha$, $\beta$ varying and $\gamma$ fixed to $0.0$ and $0.3081$ (the UPP value) -- the results are identical.}.

\begin{figure*}
	\includegraphics[width=\linewidth]{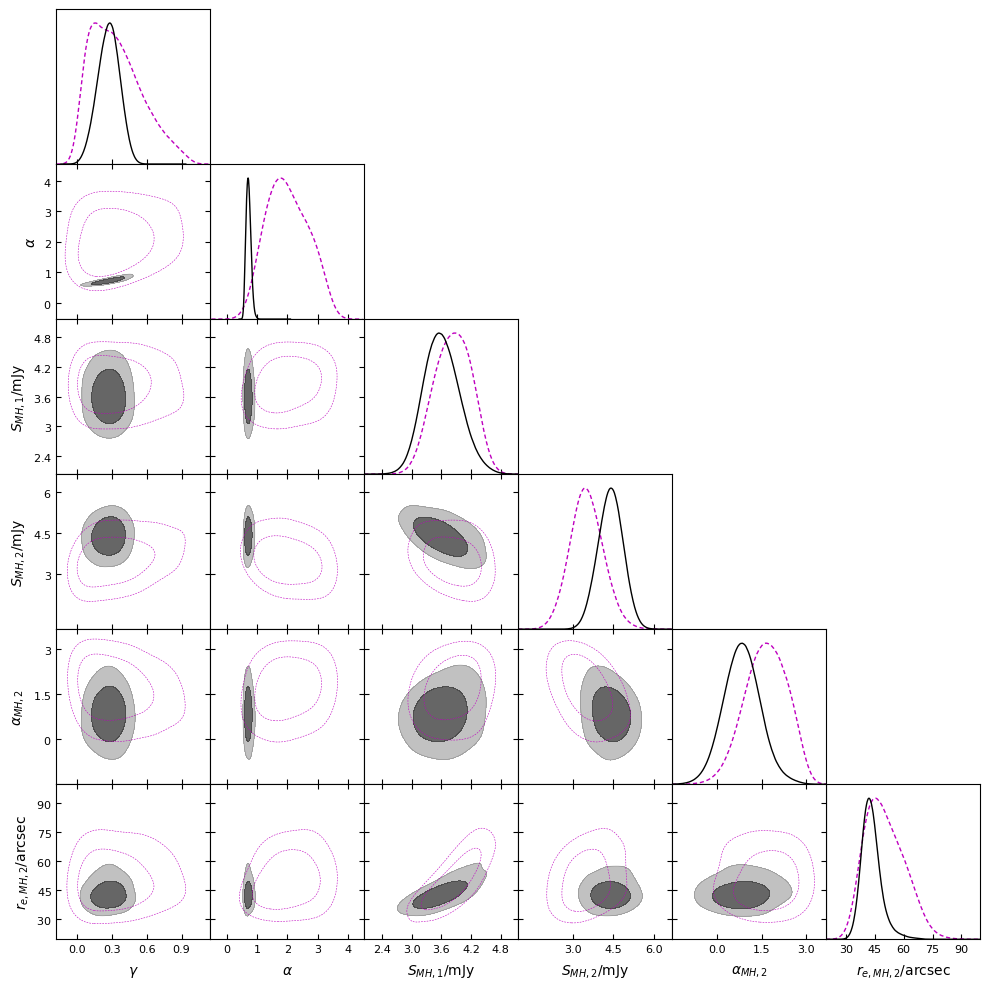} \\\vspace{-\linewidth}
     \hfill\includegraphics[width=0.35\linewidth]{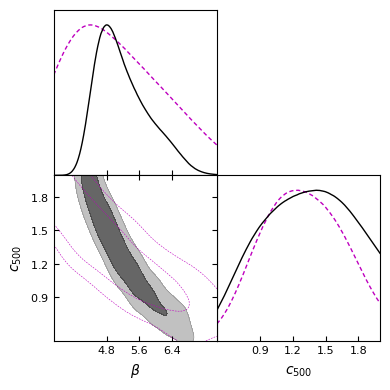}\vspace{0.7\linewidth}
    \caption{Posterior distributions of selected cluster and minihalo parameters, allowing the GNFW profile parameter shapes to vary, for \emph{Planck}$+$AMI-SA$+$LA analysis (unfilled magenta contours, dashed magenta lines) and \emph{Chandra}$+$\emph{Planck}$+$AMI-SA$+$LA analysis (filled grey contours, solid black lines).  The main plot shows the cluster and minihalo parameters which interact most when varying the profile shape parameters.  The smaller plot in the upper-right-hand corner shows the degeneracy between $\beta$ and $c_{500}$ which control the shape of the outer cluster profile; these parameters are not degenerate with the minihalo parameters.}
    \label{Fi:Chandra_SA_LA_PL}
\end{figure*}

We do not expect this to produce significant constraints on the cluster pressure profile shape since most of the information on the SZ decrement is coming from the low-resolution \emph{Planck} data, but this allows us to explore and marginalise over any degeneracy between pressure profile shape and halo flux density.  The results are reported in Table~\ref{tab:halo_fits} and posteriors for the most-affected parameters shown in Fig.~\ref{Fi:Chandra_SA_LA_PL}.  Most parameters (mass, $f_{\mathrm{gas}}$, positions) are unaffected.  The parameters of the large-angular-size halo show the most change, their posteriors showing degeneracy with the GNFW $\alpha$ parameter which controls the shape of the pressure profile at intermediate radius, however their best-fit values are all shifted by $<1\sigma$.

\begin{figure}
	\includegraphics[width=\linewidth]{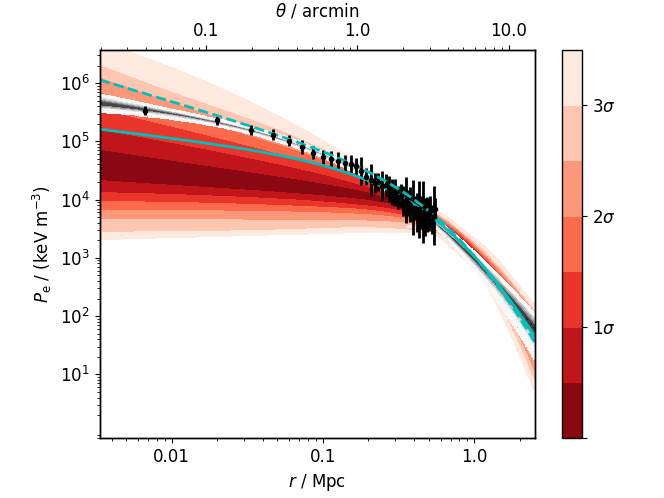}
    \caption{Constraints on the cluster pressure profile, allowing the GNFW profile parameter shapes to vary, for \emph{Planck}$+$AMI-SA$+$LA analysis (red contours) and \emph{Chandra}$+$\emph{Planck}$+$AMI-SA$+$LA analysis (grey contours).  These are plotted using \textsc{fgivenx} \citep{2019ascl.soft09014H}.  Also shown with error bars are the \emph{Chandra} deprojected pressure profile datapoints used in the analysis.  For comparison, we show in cyan the \citet{2010A&A...517A..92A} average profile for their cool-core subsample (dashed line) and UPP (solid line), normalised to the average fitted value at the characteristic radius $r_{p}$.}
    \label{Fi:pressure_profile}
\end{figure}

\begin{figure*}
	\includegraphics[trim={1cm 12.2cm 3.5cm 1cm},clip,width=0.5\linewidth]{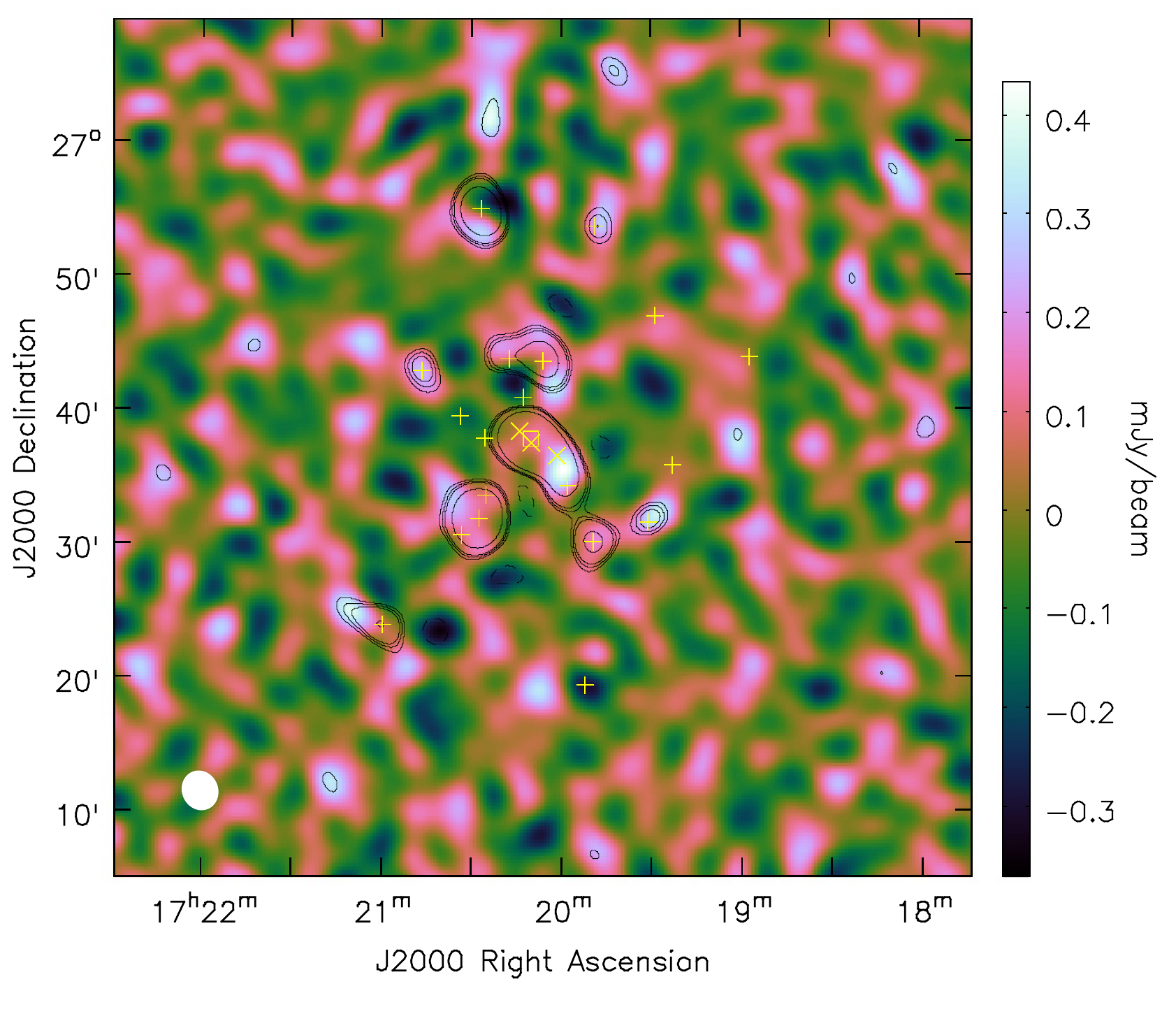}\includegraphics[trim={1cm 12.2cm 3.5cm 1cm},clip,width=0.5\linewidth]{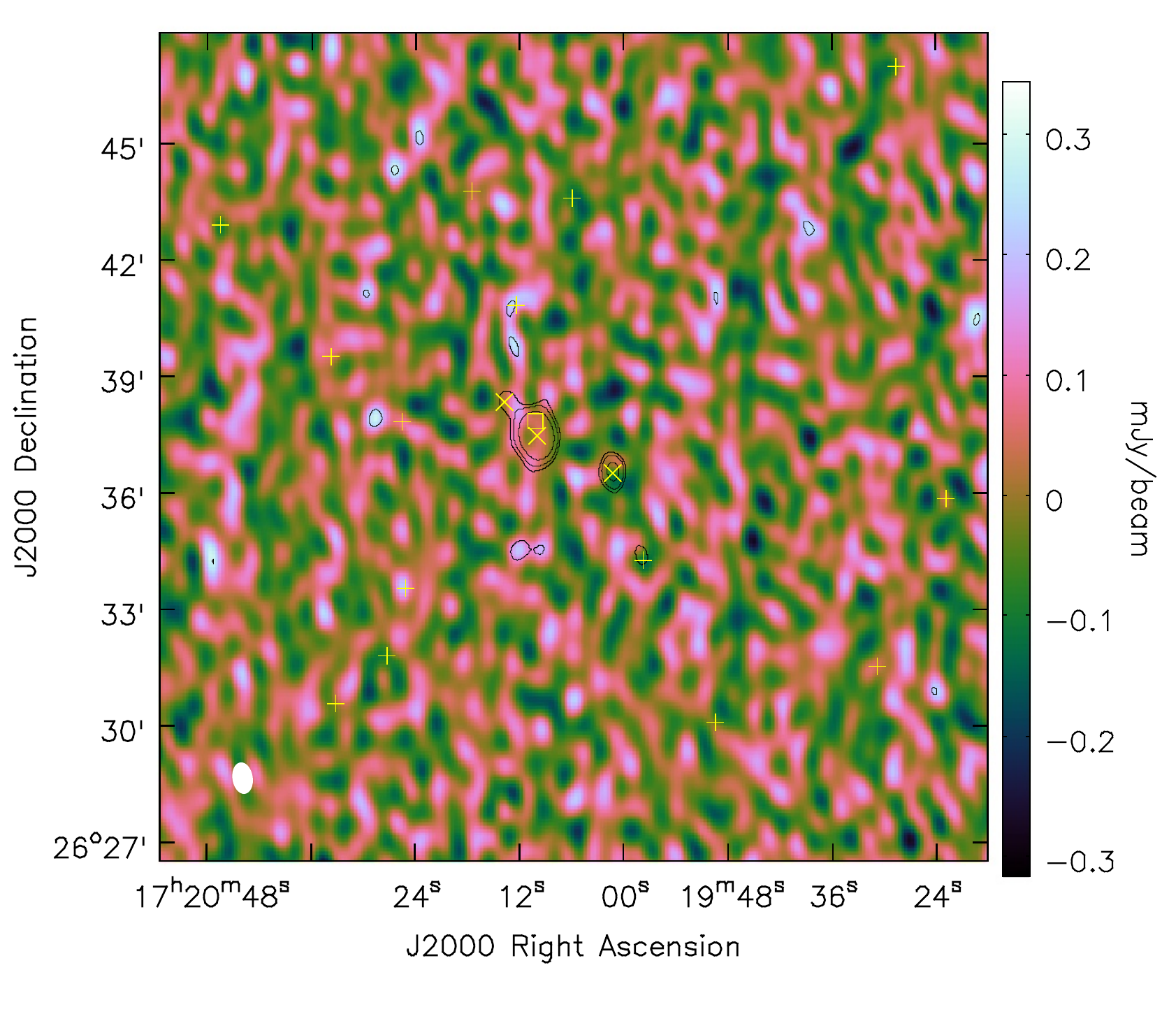}
    \caption{AMI-SA (left) and AMI-LA (right) maps after subtracting the fiducial best-fit cluster $+$ point-source $+$ double-halo model (colour-scale).  Overlaid are contours from the respective maps without any subtraction, at $\pm 3, 5, 10\sigma$ levels (negative contours dashed).  The combined model is clearly a good fit to both data sets with no significant residuals above the noise levels aside from a small under-subtraction of the source AMILA J171957+263416 on the SA, which does not affect the halo and cluster parameter fits.  The overlaid annotations show positions of point sources directly subtracted ($+$) and modelled simultaneously with the resolved sources ($\times$), and the fitted cluster position ($\square$).  The halo positions are not shown for clarity but are very close to the central, fitted source.  The solid white ellipses in the lower left-hand corners show the synthesised beams, which have dimensions 175$\times$156\,arcsec$^2$ (SA) and 47$\times$30\,arcsec$^2$ (LA).  Note that the axis scales are different as appropriate to the fields of view.}
    \label{Fi:SA_LA_sub}
\end{figure*}

To better constrain the cluster pressure profile, we make use of the deprojected pressure profile measurement for \RXJ\ from the Chandra ACCEPT archive \citep{2009ApJS..182...12C}.  We fold this into our analysis with the same priors as before.  Fitted results and posterior plots from this analysis are also shown in Table~\ref{tab:halo_fits} and Fig.~\ref{Fi:Chandra_SA_LA_PL}.  The GNFW $\gamma$ and $\alpha$ interior shape parameters are tightly constrained by this analysis, reducing the degeneracy with the halo parameters.  The inferred constraints on the cluster pressure as a function of radius, from the analysis both with and without the \emph{Chandra} data, are shown in Fig.~\ref{Fi:pressure_profile}.  We also show for comparison the \citet{2010A&A...517A..92A} average profiles for their cool-core subsample, morphologically disturbed subsample, and the `universal' pressure profile.  The fitted profile conforms most closely to the cool-core profile as expected, but is less steep in the very inner part of the cluster possibly as a consequence of the disturbance to the cool core from the merger.  Fig.~\ref{Fi:pressure_profile} shows that the X-ray and SZ data are consistent in the regions where their constraining power overlaps, giving confidence that these results are not affected by different projection biases in the two types of data.  We therefore adopt this set of results as our fiducial parameter constraints for the minihalo.  Fig.~\ref{Fi:SA_LA_sub} shows maps made from the SA and LA data with the full cluster $+$ point-source $+$ double-halo fiducial model subtracted; it is clearly a good fit to both sets of data with no significant residuals above the noise level.

\begin{figure*}
	\includegraphics[width=0.5\linewidth]{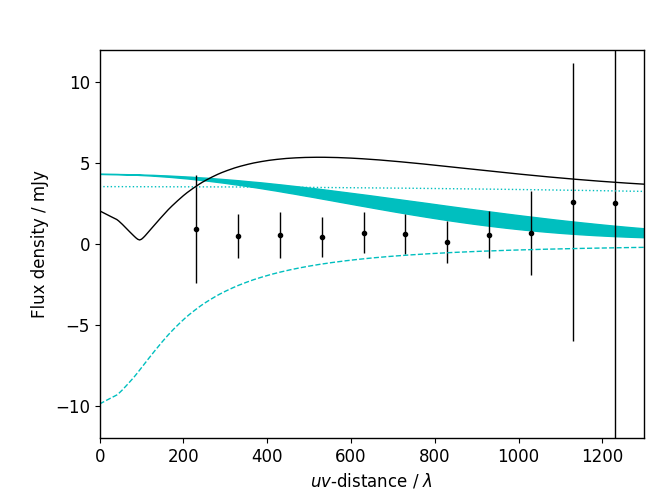}\includegraphics[width=0.5\linewidth]{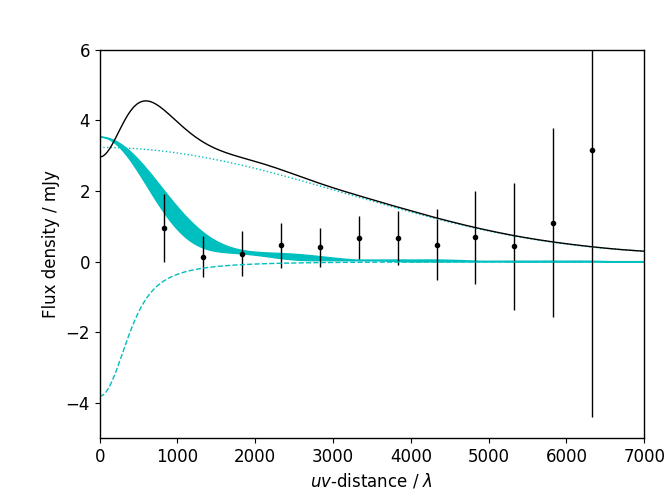}
    \caption{Here we show a representation of our final fitted model at 15.5\,GHz as a function of baseline length for the SA (left) and LA (right).  The individual components of the model are shown in cyan: the cluster model (negative) is shown with a dashed line, the smaller halo component with a dotted line, and the larger halo component with a solid band, where the width of the band shows the dependence of the model on baseline orientation.  The solid black line shows the total model.  We note that the models are different for the two arrays due to their different primary beams.  Also shown with black errorbars are the SA and LA data after subtracting all model components, averaged into bins in baseline length.  The errorbars are calculated based on the scatter of the datapoints in the bins and indicate a good agreement with the model, with no significant residuals.}
    \label{Fi:SA_LA_uvamps}
\end{figure*}

In Fig.~\ref{Fi:SA_LA_uvamps} we show the $uv$-plane models for the cluster and halo components, along with the binned residuals remaining after subtracting the total model from the data.  This illustrates the resolution effect of the interferometer and confirms that the model is a good fit to the data from both arrays at all baseline lengths.

\subsubsection{Pressure discontinuities}

\begin{figure}
	\includegraphics[width=0.8\linewidth]{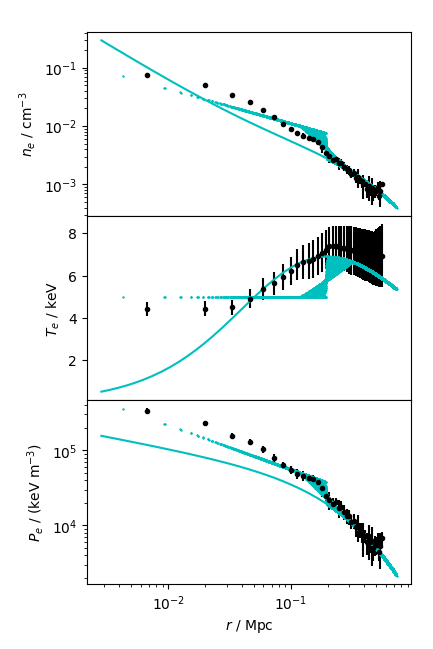}
    \caption{3D density, temperature and pressure profiles for \RXJ.  The solid cyan line in each plot shows the best-fit \emph{Planck} model with the HE cluster model and `universal' GNFW shape parameters.  The cyan points show the \citet{2001ApJ...555..205M} model for the south-eastern discontinuity combined with the \emph{Planck}-based HE model in the outskirts.  The black error bars are \emph{Chandra} deprojected estimates from the ACCEPT database.}
    \label{Fi:mazzotta_model}
\end{figure}

Although the presence of a density and temperature discontinuity (a cold front) has been demonstrated in the X-ray by \citet{2001ApJ...555..205M} and \citet{2008ApJ...675L...9M}, the density and temperature jumps go in opposite directions meaning that the pressure jump is much less significant.  Indeed, the averaged deprojected \emph{Chandra} measurements shown in Fig.~\ref{Fi:pressure_profile} (which do not take into account the cold front) show little sign of the discontinuity.  Nevertheless, we test whether it could have any effect on our measurements as follows.  We implement the discontinuity model for the more-significant south-eastern front as described in \citet{2001ApJ...555..205M}.  We replace the outer part of the model with our HE model for electron density and temperature, with mass constrained by the \emph{Planck} data, and retain the inner part (a power-law in density, and a fixed temperature) as described in \citet{2001ApJ...555..205M} and \citet{2008ApJ...675L...9M}.  The resulting radial profiles of density, temperature and pressure are shown in Fig.~\ref{Fi:mazzotta_model}.

\begin{figure}
	\includegraphics[width=0.5\linewidth]{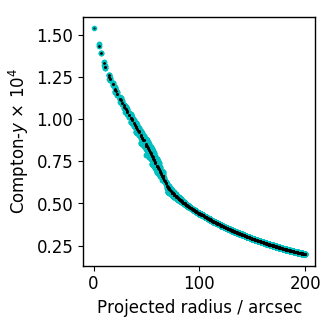}\includegraphics[width=0.5\linewidth]{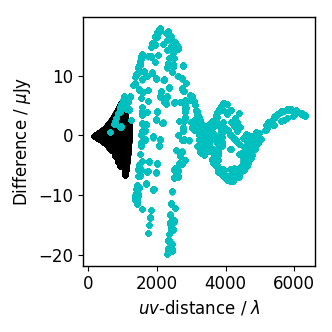}
    \caption{The left-hand plot shows simulated Compton-$y$ parameter as a function of projected radius on the sky, for the adapted \citet{2001ApJ...555..205M} model with a pressure discontinuity (large cyan dots) and the equivalent without the discontinuity (smaller black dots).  The right-hand plot shows the difference between the visibility amplitudes of noise-free simulations of these model clusters, for the AMI-SA (black dots), and AMI-LA (cyan dots), at the top end of the AMI frequency band.}
    \label{Fi:uvamp_diffs}
\end{figure}

Rather than considering the south-eastern and (less-severe) north-western discontinuities separately, we consider a model in which the cluster has two discontinuities in opposite directions consistent with the south-eastern one.  We emphasize that we are not attempting to constrain the parameters of the cold fronts here but rather assessing their maximum impact on the minihalo flux measurement.  We create a simulated Compton-$y$ parameter map for the cluster by integrating the pressure profile shown in Fig.~\ref{Fi:mazzotta_model} over the line of sight and use this as the basis for \emph{Planck}, AMI-SA and AMI-LA simulations.  We note that the line-of-sight averaging has the effect of further smoothing over the discontinuity; the Compton-$y$ profile as a function of projected radius on the sky is shown in Fig.~\ref{Fi:uvamp_diffs}.  

The cold front is at a radius of 69\,arcsec and therefore we do not expect it to affect the \emph{Planck} data which has resolution $>4$\,arcmin.  We confirm this by making mock \emph{Planck} observations of the simulated cluster with realistic noise and foreground properties consistent with the real data and analysing them with our standard \emph{Planck}-only analysis as described in Section~\ref{S:Planck-SA}.  The mass is recovered correctly whether or not we vary the GNFW profile shape parameters.  

To assess the impact on the AMI data, we create an equivalent shock-free mock cluster where the Compton $y$-parameter at a given projected radius on the sky is set equal to the average of the Compton $y$-parameter at that radius in the shocked model; this is shown in Fig.~\ref{Fi:uvamp_diffs}.  We then create noise-free simulated observations using the real $uv$-coverage from the AMI-SA and AMI-LA observations and subtract them to check the difference; a plot of this is shown in Fig.~\ref{Fi:uvamp_diffs} for the top of the AMI frequency band (maximum SZ signal).  The difference is $<10$\,$\upmu$Jy for the SA and $<20$\,$\upmu$Jy for the LA and therefore insignificant compared to both the noise levels in the data and the flux density of the halo.  We also add (single) minihalos with parameters similar to those preferred by the SA and LA data to the cluster model, and confirm that the halo parameters are recovered correctly from joint \emph{Planck}-SA-LA analysis of simulations with appropriate noise levels.  We can therefore be confident that the pressure discontinuities are not affecting our halo parameter constraints.

\subsubsection{BCG core}

\begin{figure}
	\includegraphics[width=0.575\linewidth]{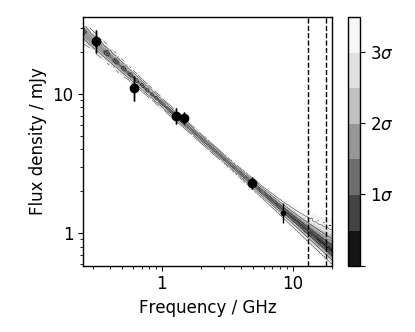}\includegraphics[width=0.425\linewidth]{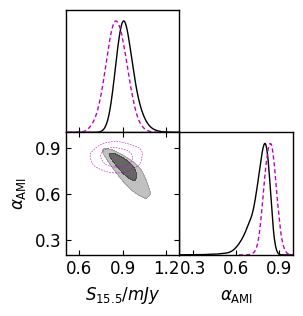}
    \caption{BCG spectrum, fitted using a two-component model to investigate the possibility of a flat-spectrum core (see text for more details).  The left-hand plot shows the datapoints from G14 and the fitted spectrum; the right-hand plot shows the inferred constraints on the 15.5\,GHz flux density and the AMI in-band spectral index (filled constraints, solid black lines) compared to the single-component constraints (unfilled magenta constraints, magenta dashed lines).}
    \label{Fi:flat_core}
\end{figure}

Since the AMI data does not have the angular resolution to separate the point-like BCG from the extended radio emission, we have relied on an extrapolation of the power law fit between 317\,MHz and 8.44\,GHz using the datapoints presented in G14.  The datapoints are extremely consistent with the fit despite being taken over a wide range in time from 1985 -- 2009, so we have confidence that the extrapolation will not be affected by variability.  However, faint ($<1$\,mJy), compact radio galaxies have been shown to often contain a flatter-spectrum core which becomes increasingly important at 15\,GHz (e.g.\ \citealt{2020MNRAS.493.2841W}).  Similarly, \citet{2015MNRAS.453.1201H} investigated a large sample of radio BCGs and found that many could be spectrally decomposed into a steep-spectrum ($\alpha>0.5$) and flat-spectrum ($\alpha<0.5$) component.  We investigate the possibility that the \RXJ\ BCG contains a flat-spectrum core as follows.  Firstly, we fit the datapoints between 317\,MHz and 8.44\,GHz using a two-component model:
\begin{equation}
S(\nu) = S_{0,s} \left ( \frac{\nu}{\nu_0} \right )^{-\alpha_{s}} + S_{0,f} \left ( \frac{\nu}{\nu_0} \right )^{-\alpha_{f}}
\end{equation}
\noindent where the reference frequency $\nu_0 = 8.44$\,GHz.  For the steep-spectrum component, we use the fit to frequencies $<8$\,GHz as a starting point.  We set the prior on $S_{0,s}$ as uniform in the range $S_0 \pm 5\upDelta S_0$ from the low-frequency fit, and set the prior on $\alpha_{s}$ as $\mathcal{U}[0.5, 2.0]$ to avoid mode-swapping.  For the flat-spectrum component, we give $S_{0,f}$ a wide prior, $\mathcal{U}[0, 10]$\,mJy and use the \citet{2015MNRAS.453.1201H} distribution of core spectral indices as the prior on $\alpha_{f}$.  The resulting fitted total spectrum is shown in Fig.~\ref{Fi:flat_core}; although the lower-frequency points tightly constrain the steep-spectrum component, the error bar on the 8.44\,GHz point allows for a flatter component to be present at higher frequency.

We use this fit to constrain the 15.5\,GHz flux density and AMI in-band spectral index, by fitting a power law across the AMI band to the sum of the two model components.  The flux density and spectral index are correlated as shown in Fig.~\ref{Fi:flat_core}, i.e.\ a higher flux density is allowed if the total spectrum is flatter.  We check whether the presence of a core could impact our minihalo parameter constraints by using this two-dimensional posterior constraint as a prior on the BCG flux density and spectral index; we draw samples from the posterior and add an extra random variation using Gaussian priors with width 5 per cent and 0.03 in flux density and spectral index respectively to account for possible calibration error.  To save computational time we omit the \emph{Planck} data, instead using a Gaussian prior based on the fiducial constraint on $M_{200}$; fix $\beta$ and $c_{500}$ to their values from the fiducial analysis, and fit a single-halo model to (i) the \emph{Chandra}$+$LA dataset and (ii) the \emph{Chandra}+SA dataset.  In both cases we see no significant change to the posterior on the minihalo parameters.  We can understand this by comparing the one-dimensional priors on the flux density as shown in Fig.~\ref{Fi:flat_core}.  The shift is very small and is negligible compared to the noise on the data and other model-fitting uncertainties, so the posteriors on the minihalo parameters are unchanged.  Likewise, although the shift in spectral index seems larger, its effect over the limited frequency range of the AMI band is negligible compared to the other uncertainties.  We can therefore be confident that our minihalo results are robust to the presence of a flat core in the BCG.
 
\subsection{Comparison to other frequencies}

G14 analysed radio data between 317\,MHz and 8.44\,GHz to construct power-law spectra of the central part and tail of the minihalo.  We only consider the central part here since the tail, which is both fainter and has a steeper spectrum, is not detected in the AMI observations.  G14 used a map-based method to estimate flux densities, enforcing a common range of baseline lengths in the $uv$-plane of 1 -- 50\,k$\lambda$ and imaging with natural and/or uniform weights.  They noted that this requires high sensitivity and good sampling of the short baselines to recover the extended, low-surface-brightness emission of the minihalo.  This method gave flux densities following a consistent power-law with $\alpha=1.0\pm0.1$ between 317\,MHz and 4.86\,GHz, however the 8.44\,GHz flux density was lower.  G14 suggested this may indicate a high-frequency steepening of the spectrum, with the caveat that the $uv$-coverage at 8.44\,GHz had much poorer sampling of the smallest $uv$-distances compared to the other data.

The AMI data we are considering have necessitated a different approach.  The LA data cover $\approx$\,600 -- 6000$\lambda$, but the sampling between $\approx$\,1 -- 2\,k$\lambda$ is too poor to reliably recover the extended flux density.  The SA data fill in this region in $uv$-space, but mostly cover much shorter $uv$-distances so making a map with baselines $> 1$\,k$\lambda$ length would be challenging.  To reliably combine the two datasets (plus disentangle the SZ emission), the $uv$-plane fitting approach was more appropriate.  To properly compare our flux estimate with the others at lower frequency, we need to calculate how much emission would be seen on a map at AMI frequencies made from a visibility dataset with dense, uniform $uv$-sampling down to 1\,k$\lambda$.  To do this, we construct our double-halo model, Fourier transform to the aperture plane, and take the value of the amplitude at 1\,k$\lambda$.  This gives a flux density loss of 35 per cent -- 45 per cent for the best-fit halo values depending on baseline orientation.  Assuming dense $uv$-coverage this means the flux density measured on the map should be the total flux density for the model multiplied by 0.65 (i.e.\ the minimum flux loss because the dense sampling should ensure the baselines in the correct orientation to sample the most flux are included).  We use \textsc{fgivenx} \citep{2019ascl.soft09014H} to sample our posteriors for the halo parameters and calculate the maximum aperture plane flux density at 1\,k$\lambda$ as a function of frequency for each sample in the posterior, giving an overall constraint on the spectrum between 13 -- 18\,GHz as shown in Fig.~\ref{Fi:final_spec}.  Our constraint is entirely consistent with the extrapolation from lower frequency with no sign of steepening.  For comparison, we also include datapoints where we have taken the simpler approach of subtracting the best-fit cluster $+$ radio-point-source model and mapping the LA and SA data; the LA using baselines $>1$\,k$\lambda$ length and the SA using all data.  The two datapoints bracket the $uv$-plane results showing the importance of accounting for $uv$-plane sampling correctly for this resolved source.

\subsubsection{8.44\,GHz VLA data}

We now test whether a $uv$-plane analysis similar to the approach we have used for the AMI data reconciles the apparent difference between the VLA datapoint at 8.44\,GHz and the rest of the measured spectrum, using the same VLA dataset as G14 (Project AH0355, DnC configuration, 1989 Jun 2).  We note that this is an $\approx$\,3\,minute observation meaning the $uv$-coverage is sparse and there are few short baselines present in the data.  The SZ effect is much less significant at this frequency with a maximum amplitude of 80\,$\upmu$Jy on the shortest baseline according to the best-fit model from the \emph{Chandra}$+$\emph{Planck}$+$SA$+$LA analysis, so we merely subtract the best-fit cluster model rather than marginalizing over cluster parameters in this analysis.  Only three of the radio sources appear in the VLA map -- the BCG, the head--tail galaxy, and AMILA\,172001+263632; we take their positions from the VLA map (in the case of the BCG, from a map made with baselines $>15$\,k$\lambda$ length to exclude the minihalo emission) and fit their flux densities simultaneously with the minihalo parameters using \textsc{McAdam}.  The VLA dataset has only one frequency so we do not fit any spectral indices.  We use the same external priors on the flux density of the BCG and the head--tail galaxy as in the AMI analysis and do not find any disagreement between the data and the priors.  In the case of AMILA\,172001+263632 we use a Gaussian prior centred on the flux density measured from the map with a width of 50 per cent.

\begin{figure*}
	\raisebox{-0.5\height}{\includegraphics[trim={1cm 12.2cm 3.5cm 1.5cm},clip,width=0.5\linewidth]{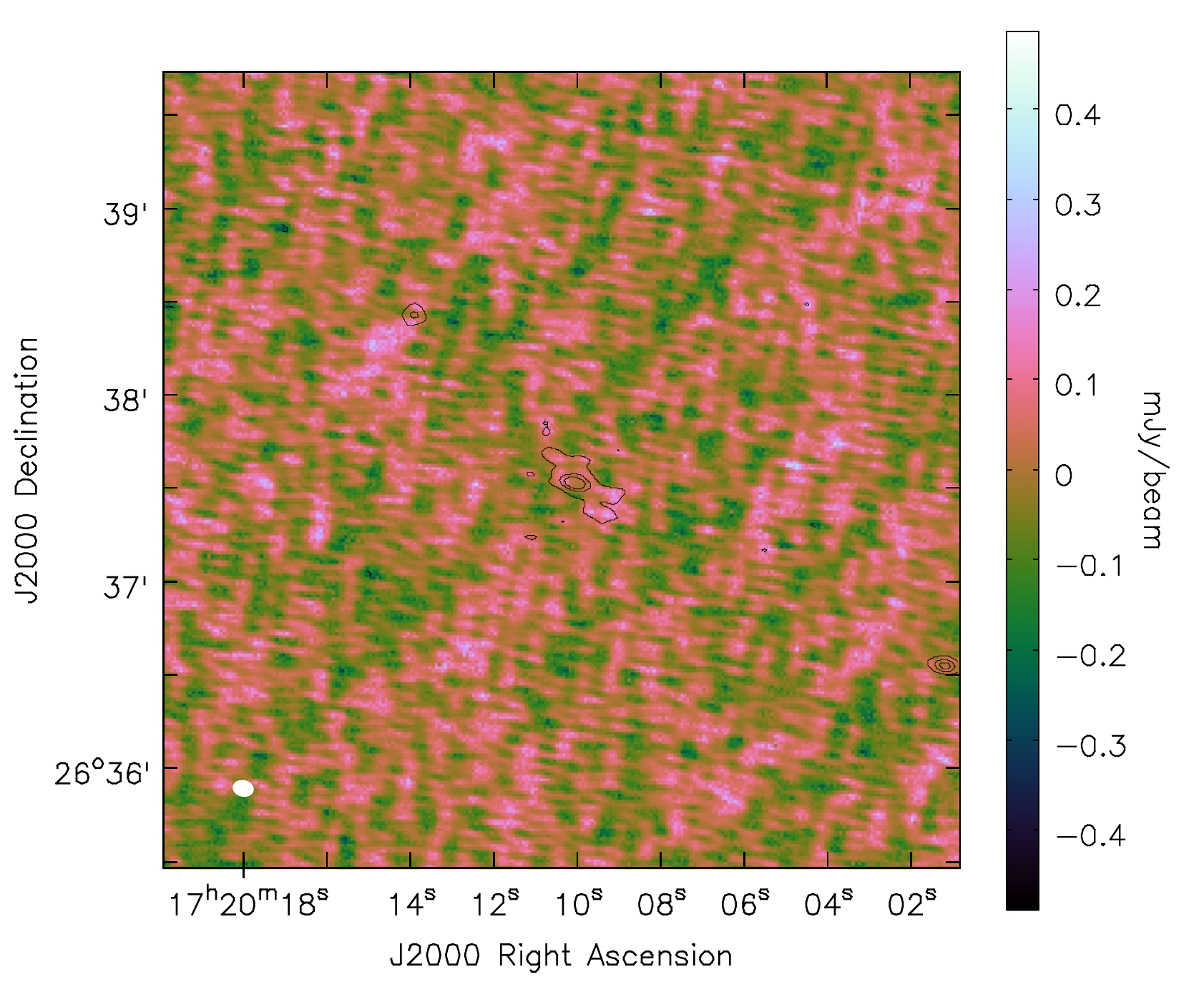}}\raisebox{-0.5\height}{\includegraphics[width=0.5\linewidth]{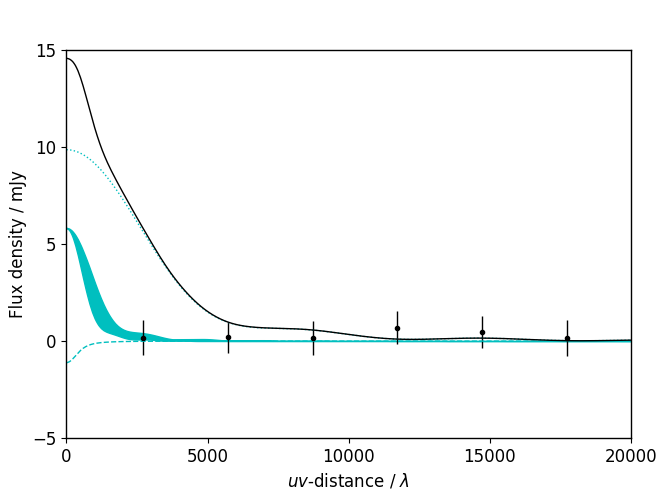}}
    \caption{Left: map of VLA 8.44\,GHz data after subtracting the best-fit single-halo, point source and cluster model (colour-scale) with contours from the unsubtracted map overlaid starting at $3\sigma$.  There are no residuals at $>3\sigma$ level in the residual map.  The synthesised beam is shown as the white solid ellipse in the lower left-hand corner and has dimensions 6.5$\times$5.0\,arcsec$^2$.  Right: as in Fig.~\ref{Fi:SA_LA_uvamps}, best-fit model visibility amplitudes for the cluster (dashed cyan line), smaller halo (dotted cyan line) and larger halo (solid cyan band) components and total model (black line).  The black points with error bars are the data points, binned in $uv$-space after subtracting the total model, showing no significant residuals.  The data extends to baseline length $\approx$\,60\,k$\lambda$ but we have excluded the longer baselines where the extended signal is negligible.}
    \label{Fi:VLA}
\end{figure*}

As in the case of the AMI data, we initially fit a single exponential minihalo model using the priors listed in Table~\ref{tab:halo_priors}.  The fitted parameters are listed in Table~\ref{tab:halo_fits}; the size and shape of the halo agree well with the AMI-LA single-halo constraints, which are most comparable in terms of minimum baseline length.  We next attempted to fit the double-halo model with the wide, non-informative priors used in the AMI analysis.  Although there was a slight preference for the double-halo model with $\log(\mathcal{Z}_{\mathrm{double}}/\mathcal{Z_{\mathrm{single}}}) = 1.8 \pm 0.3$, there was not enough information in the data to constrain the parameters of the larger halo.  Even when applying a Gaussian prior based on the AMI constraints on $r_{e,2}$ the parameters of the larger halo were entirely unconstrained; we therefore applied Gaussian priors also on $S_{\mathrm{minihalo},2}(8.44\,\mathrm{GHz})$ based on the AMI constraints on $S_{\mathrm{minihalo},2}(15.5\,\mathrm{GHz})$ and $\alpha_{\mathrm{minihalo},2}$.  This again gave a preference for the double-halo model with $\log(\mathcal{Z}_{\mathrm{double}}/\mathcal{Z_{\mathrm{single}}}) = 2.8 \pm 0.3$ (considered `decisive' according to \citealt{jeffreys}), but comparing the best-fit model visibilities we see that they are almost indistinguishable for the single- and double-halo cases on the baselines present (see Fig.~\ref{Fi:VLA}) so a dataset with more short baselines would be required to confirm whether the double-halo shape persists at frequencies lower than 15\,GHz.  Both models subtract cleanly from the data; there are no residuals above the noise level in the subtracted maps as shown in Fig.~\ref{Fi:VLA}.

We use the same method as described above for the AMI data to extract constraints on the flux density at baseline length $>1$\,k$\lambda$; this adjusted flux density point is shown in Fig.~\ref{Fi:final_spec} for both the double- and single-halo model.  Both agree well with the spectrum extrapolated from lower frequency and it is clear that the apparent steepening of the spectrum was caused by incomplete $uv$-sampling.

\subsubsection{28.5\,GHz BIMA measurement}

\begin{figure}
	\includegraphics[width=\linewidth]{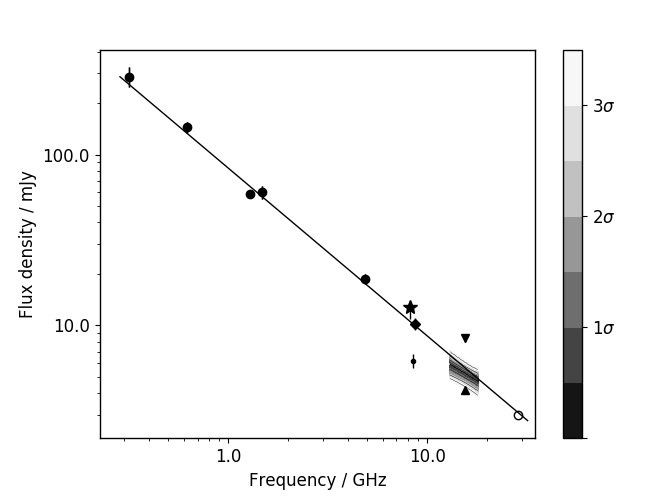}
    \caption{Spectral energy distribution of the central part of the minihalo.  The points marked with large circles and error bars are taken from G14 and are used to fit the power-law spectrum shown with the solid line and extrapolated over the full frequency range.  The small circle with error bar below the fit is the 8.44\,GHz VLA measurement from G14.  The star and diamond markers show the VLA double- and single-halo $uv$-plane-fitted estimates respectively, displaced horizontally for clarity.  The $1-3\sigma$ constraints shown in greyscale are the results of the $uv$-plane fitting to the AMI data; the triangles below and above are map-based measurements using the LA and SA data respectively showing the effect of the different $uv$-sampling.  The open circle is the 28.5\,GHz measurement from \citet{2007AJ....134..897C}; it is a rough estimate only as the SZ decrement has not been removed and the $uv$-sampling is not accounted for.}
    \label{Fi:final_spec}
\end{figure}

\citet{2007AJ....134..897C} observed a sample of clusters including \RXJ\ with the Berkeley-Illinois-Maryland Association (BIMA) array at 28.5\,GHz and investigated their radio source environments.  They detected no SZ decrement toward the source but reported a radio source with radius 36.6\,arcsec and flux density 2.99$\pm$0.17\,mJy coincident with the minihalo position.  They do not give the $uv$-coverage of their observations but note that the telescopes were configured in a compact configuration providing dense $uv$-coverage to the shadowing limit.  Their minimum baseline length should therefore be given by the telescope diameter of 6.1\,m, i.e.\ $\approx$\,580\,$\lambda$ at 28.5\,GHz.  Their datapoint will be biased down by the presence of the SZ decrement, but biased up in comparison to the datapoints at other frequencies since it includes baselines $<1$\,k$\lambda$.  We plot the point without error bar in Fig.~\ref{Fi:final_spec}; although we cannot account for the SZ and $uv$-sampling without access to the data it is remarkably consistent with the extrapolation from lower frequency and provides further evidence that the spectrum does not steepen as inferred from the 8.44\,GHz datapoint by G14.

\section{Discussion}\label{S:discussion}

Our high-frequency radio observations provide a unique testing-ground for discriminating between competing models for the production of the relativistic electrons required to produce the synchrotron emission seen in minihalos.  The `re-acceleration' model proposes that electrons injected into the ICM by the central AGN are accelerated to relativistic energies by turbulence caused by `sloshing' motions of the cool core of the cluster, triggered by gravitational interactions with subclusters \citep{2002A&A...386..456G}.  Alternatively, the `hadronic' or `secondary' model proposes that thermal protons throughout the cluster volume are accelerated to relativistic energies by supernovae, AGN and ICM shocks; these cosmic ray protons undergo interactions with the thermal proton population, producing pions which decay into a series of products including relativistic electrons (e.g.\ \citealt{2004A&A...413...17P}).

\citet{2013ApJ...762...78Z} and \citet{2015ApJ...801..146Z} (hereafter Z13 and Z15) investigate the observational consequences of these two production mechanisms using magneto-hydrodynamical simulations of a similar system to \RXJ, i.e.\ a massive cool-core galaxy cluster perturbed by the passage of a smaller subcluster.  This causes sloshing, turbulence and magnetic field amplification in the core region and produces X-ray cold fronts similar to those observed in \RXJ.  The two different electron production methods result in different predictions for the synchrotron emission, both in terms of its spectrum and spatial structure.  Here we outline these differences and compare to our observations.

\subsection{Spatial structure}\label{S:spatial_extent}

In the `re-acceleration' simulation, Z13 show that the synchrotron emission is tightly confined within the cold front region where the turbulence is greatest.  If electrons are injected into bubbles blown by the AGN in the ICM, they are redistributed as they are accelerated by the sloshing, resulting in a more diffuse distribution inside the cold fronts, in agreement with many (low-frequency) observations of minihalos; this has often been interpreted as support for the re-acceleration scenario (e.g.\ \citealt{2008ApJ...675L...9M}, \citealt{2013ApJ...777..163H}, G14).

Z15 show that in the `hadronic' scenario, in contrast, the emission is not so tightly constrained within the cold fronts.  The relativistic protons have long radiative loss times compared with the age of the cluster, so diffuse further through the cluster interacting with thermal protons throughout the ICM and producing the required relativistic electrons.  The synchrotron emissivity depends on the magnetic field strength which is amplified by the turbulence inside the cold fronts, so the emission is strongest inside the cold fronts as in the re-acceleration case, but weaker emission is also seen outside the cold fronts.

\begin{figure}
	\includegraphics[trim={1cm 13.5cm 5cm 2.2cm},clip,width=\linewidth]{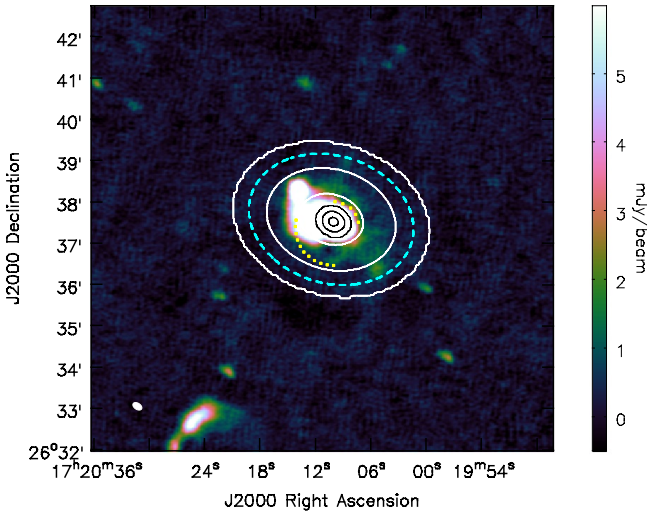}
    \caption{Low-resolution 144\,MHz LOFAR map of the cluster field (colour-scale), showing the central part of the halo.  The colour-scale is truncated to display the low-surface-brightness features.  Overlaid in black and white respectively are the contours of the fitted double-halo model at 1, 2 and $3\times r_{e}$ for the small and large halo.  The thick, dashed cyan contours show the boundary of the 15.5\,GHz emission at $2.5\times r_{e,2}$ as determined using the AMI-SA data.  The yellow squares show the positions of the cold fronts as determined by \citet{2008ApJ...675L...9M}.  The LOFAR synthesised beam is shown in the bottom left-hand corner as the solid ellipse.}
    \label{Fi:LOFAR_halos}
\end{figure}

In particular for \RXJ, G14 show that the minihalo emission as observed at 617\,MHz by the GMRT is confined by the cold fronts detected in X-ray, extending to around 80\,kpc radius from the cluster centre.  The baseline length of these observations goes down to $\approx$\,220\,$\lambda$, similar to our AMI-SA observations, so the lack of emission seen by the GMRT at larger scales should not be due to incomplete $uv$-coverage.  The dimensions of our high-frequency smaller halo agree with this; however the larger halo extends far beyond the cold fronts.  We test how far out our data actually require the halo to extend by cutting off our best-fit model at 1, 2, 2.5, 3 and 4 $\times r_{e,2}$ (in the model-fitting it is cut off at 3$\times r_{e,2}$), Fourier-transforming and subtracting the model visibilities from the real AMI-SA data, mapping the result and checking the significance of the residuals.  When we cut off the model at $2.5\times r_{e,2}$ we see a faint residual at $\approx 1\sigma$ at the minihalo position; when we cut off at $3\times r_{e,2}$ we see no residual above the noise level.  We therefore conclude that our data require the minihalo to extend from the centre to at least $\approx 2.5 \times 43.1 / \sqrt{f} = 124$\,arcsec along the major axis, or $\approx$\,340\,kpc at the redshift of the cluster giving it a total extent of 680\,kpc along the major axis.  Fig.~\ref{Fi:LOFAR_halos} shows the sizes and locations of the fitted halos in comparison to a LOFAR map of the cluster field and the X-ray cold fronts; it is clear that the smaller halo is confined by the cold fronts while the larger halo extends far past them.  The ellipticity of the larger halo does however seem to indicate some correlation with the cold front structure since the halo extends further in the direction parallel to the cold fronts than perpendicular to them.

\citet{2019A&A...622A..24S} use LOFAR observations of the cluster at 144\,MHz to also demonstrate the existence of a larger-scale halo in addition to the minihalo.  They do not detect this larger-scale halo in their reprocessed 617\,MHz GMRT map with $uv$-coverage matched to the LOFAR map, showing that it must have an ultra-steep spectrum.  We would therefore not expect to detect it at 15.5\,GHz.  Fig.~\ref{Fi:LOFAR_halos} shows the LOFAR low-resolution image; the large-scale halo detected by AMI extends even further than the halo detected by LOFAR.  The LOFAR configuration provides dense $uv$-coverage with sensitivity up to degree scales \citep{2017A&A...598A.104S} so the lack of emission observed by LOFAR on the scale of the AMI halo should not be a resolution effect.  We therefore consider the ultra-steep spectrum halo to be unrelated to the high-frequency halo.

Z15 note that \citet{2013MNRAS.434.2209W} discuss momentum-dependent diffusion affecting the spectral distribution of the relativistic protons.  Although this effect is not included in their simulations they predict it would result in a spectral \emph{flattening} of the synchrotron emission with distance from the cluster centre, since higher-energy protons would travel further through the ICM.  This would result in minihalos appearing broader at higher frequencies, consistent with our observations.  The constraint on the spectral index over the AMI band ($\alpha_{\mathrm{MH},2}=0.84 \pm 0.62$) is too weak to test this prediction due to the limited frequency lever-arm of the AMI observations.  We note that the VLA data used by G14 at frequencies above 1\,GHz has much longer minimum baseline lengths, with $uv > 500\lambda$.  The larger halo could therefore be present at these frequencies but undetected by the observations due to the lack of short baselines; more observations at intermediate frequencies with short baselines are required to test whether the large-scale emission detected by AMI can be explained by the flattening phenomenon.  More simulations would also be required to test whether a similar effect could also be caused by multiple AGN outbursts and/or multiple cluster merger events.

\subsection{Emission spectrum of the minihalo}

In the `re-acceleration' scenario, the energy balance between re-acceleration and energy losses through synchrotron and inverse Compton radiation naturally creates high-frequency steepening of the radio spectrum, with the frequency at which this occurs dependent on the acceleration efficiency.  Z13 make time-dependent predictions for the integrated emission spectrum of the minihalo, which range from $\alpha \approx 1.5$ -- 2.0 and begin to steepen above $\approx$\,1\,GHz (their fig.~17), with time-dependent slopes.

In the `hadronic' scenario, the proton--proton interactions continually refresh the population of relativistic electrons.  Under relatively stable environmental conditions, a steady state is reached where the energy gains and losses balance out and the radio spectrum is a simple power law, the spectral index of which depends only on the energy spectrum of the relativistic electrons.  Z15 show that the turbulence generated by the merger does not disrupt this scenario enough to change the integrated spectra, which follow a single power law over a wide range in frequency with $\alpha \approx$\,1.1 (their fig.~8).

The integrated emission spectrum of the central part of the \RXJ\ minihalo which we are considering here is remarkably consistent with a single power law with $\alpha = 1.0$ over an extremely wide range of frequencies, from 300\,MHz to 18\,GHz (the top of the AMI band) and potentially even up to 28.5\,GHz.  We note that since the flux densities in the spectrum shown in Fig.~\ref{Fi:final_spec} are at $uv>1$\,k$\lambda$ they are mostly probing the smaller halo, and the potential spectral flattening of the larger-scale halo mentioned in Section~\ref{S:spatial_extent} will be unimportant.

The lack of spectral steepening combined with the less-steep $\alpha = 1.0$ spectral index seem to support the hadronic scenario.  However, the Z13 simulations are based on one initial injection of electrons rather than a continuous injection as would be expected if the BCG has an AGN.  If the AGN is injecting enough energy to balance out the synchrotron and inverse Compton losses, there would be no spectral steepening.  G14 argue that the \RXJ\ BCG may not be active because it is weak ($\approx$\,5$\times 10^{23}$\,W\,Hz$^{-1}$ at 1.48\,GHz) and compact ($<1.4$\,kpc) with no radio lobes or jets visible, nor X-ray cavities visible in the ICM.  However they also argue that since its spectral index is relatively steep ($\alpha\approx$\,0.8), similar to other extended, active radio galaxies, it may in fact be active with radio jets/lobes existing on sub-kiloparsec scales.  

We note that \citet{2012MNRAS.423..422L} classify the BCG as a `combination' galaxy (star-forming and AGN) based on optical line strength ratios, and \citet{2016MNRAS.461..560G} classify it as `active' (with either star-forming or AGN activity) on the basis of its optical and near-UV colours.  Inspection of the SDSS DR14 \citep{2018ApJS..235...42A} spectrum (plate 979, fiber 141, MJD 52427; aperture 2.0\,arcsec diameter) of the BCG shows it to be of low-ionisation nuclear emission-region (LINER) type \citep{1980A&A....87..152H}, in particular because of its relatively strong [SII] 6716, 6730\AA\ forbidden emission lines.  The optical spectrum bears a distinct resemblance to that of PKS\,2322$-$12 (\citealt{1977ApJ...211..675C}; $z = 0.082$, aperture $2.7 \times 4$\,arcsec$^{2}$, resolution 7\,\AA) which is also a giant cD galaxy and the BCG of a cool-core cluster, Abell\,2597.  The central radio source of PKS\,2322$-$12 has radio, milliarcsec, oppositely directed jets \citep{1999ApJ...512L..27T} and spatial and spectral evidence of jets powering its radio emission on a scale of a few kpc \citep{2005MNRAS.359.1229P}.

We next test, {\em assuming} the \RXJ\ BCG to be powered by jets, whether its total near-UV $+$ optical narrow-line (forbidden + recombination) luminosity $L_{\rm NL}$ places it amongst other jet-powered radio sources. We use the jet bulk power $Q$ versus $L_{\rm NL}$ correlation from \citet{1991Natur.349..138R} (RS91) which takes $L_{\rm NL} = 9 L_{[{\rm OII}]3727} + 4.5 L_{[{\rm OIII}]5007}$ as a proxy. We use this particular correlation for two reasons.
\begin{itemize}
\item The $Q$ needed to run a radio source is evidenced by the energy of the radio source in magnetic field and relativistic particles.  The equipartition energy is very close to the {\em minimum} energy required to produce the observed synchrotron emission (\citealt{1959ApJ...129..849B}, and see \citealt{1980ARA&A..18..165M}), plus the work done in expanding against the intergalactic medium, divided by the age of the radio source.  We note that radio luminosity is no proxy for $Q$.
\item Most radio sources must contain a ``machine'' (now widely accepted to be a massive, rapidly spinning black hole plus an accretion system) that produces a pair of jets.  Internal energy from accretion (for which narrow lines are a major mechanism, radiating energy out of the system and promoting accretion) is converted to bulk energy in the jets.
\end{itemize}

From the SDSS spectrum, we estimate $L_{\rm NL} = 2.2 \times 10^{35}$\,W for $H_0 = 50$\,km\,s$^{-1}$\,Mpc$^{-1}$ as assumed in RS91, which puts it in the transition area between Fanaroff-Riley (FR)\,1 and FR\,2 radio structure \citet{1974MNRAS.167P..31F} on the RS91 graph as expected if the \RXJ\ BCG is powered by two jets.

Finally, \citet{2021MNRAS.500.4749B} have observed a sample of nearby galaxies at high angular resolution with the e-MERLIN telescope.  They find that the elliptical galaxies with LINER-type spectra in their sample tend to have jets and FR\,1-like core-brightened radio morphologies.

Taken together, these pieces of evidence suggest that the \RXJ\ BCG is probably active and producing radio jets which would explain the lack of spectral steepening in the context of the `re-acceleration' scenario for the minihalo (but not the larger spatial extent at high frequency).  We plan to observe it with e-MERLIN with 0.05\,arcsec resolution at C-band to check for the presence of sub-kiloparsec jets; the difficulty in detecting such jets lies in matching the interferometer-array response to a possible pair of jets, each of which is of coherent but unknown 3D structure, seen in projection on the sky.

\subsection{Spatial variation of the spectrum}

We have not considered spatial variation of the minihalo spectrum in our modelling of the high-frequency halo due to lack of resolution and sensitivity.  However we note that \citet{2019A&A...622A..24S} show a spectral index map between 144 and 610\,MHz, showing that the central part of the halo is consistent with a single spectral index of $\approx$\,1.0 (their fig.~10) while G14 show a spectral index map between 617\,MHz and 1.48\,MHz, also showing a relatively constant spectral index at the centre with spectral steepening up to $\alpha \approx$\,2.5 along the `tail' of the minihalo which follows the spiral cold front structure (their fig.~8).  They show that this steepening cannot be explained by ageing of relativistic electrons which are advected or diffuse to its periphery since the velocities required for these propagation mechanisms are too large.  Z15 simulations do show a similar spectral steepening in the `hadronic' case along the spiral due to rapid magnetic field amplification, but only up to $\alpha=1.3$; the observed steepening is more consistent with the re-acceleration model where the acceleration efficiency depends critically on the distribution of the turbulence.  It may be therefore that \emph{both} mechanisms contribute to the production of relativistic electrons, with turbulent re-acceleration more important in the inner part of the minihalo observed at all frequencies, and hadronic production more important in the outer part as observed at 15\,GHz.

\section{Conclusions}\label{S:conclusions}

We have presented observations of the \RXJ\ minihalo at 13 -- 18\,GHz, which is the highest-frequency detection of a minihalo to date.
\begin{enumerate}
\item{We have successfully and robustly disentangled the SZ and minihalo emission, which exist on the same angular scales, using ancillary SZ and X-ray data from \emph{Planck} and \emph{Chandra}.}
\item{By model-fitting in the $uv$-plane we have combined interferometric datasets with different $uv$-ranges and constrained a model for the minihalo emission.}
\item{The minihalo emission at this frequency is not well-described by the single exponential-halo model which has been successful for other minihalos at lower frequency, but requires a double exponential-halo model.}
\item{The smaller exponential halo in the fitted model agrees well with data at lower frequencies where the minihalo emission is confined within cold fronts in the ICM detected in X-ray.}
\item{The larger exponential halo in the fitted model extends well beyond the cold fronts, and has an ellipsoidal shape directed parallel to the cold fronts.}
\item{The 13 -- 18\,GHz measurements are entirely consistent with an extrapolation of the spectrum from lower frequency with $\alpha = 1.0$; reanalysis of a VLA dataset at 8.4\,GHz using our $uv$-plane modelling method shows that previously reported indications of spectral steepening were due to interferometric filtering of large angular scales.}
\item{By comparison with magneto-hydrodynamical simulations of a similar cluster, we show that our high-frequency observations are more consistent with the `hadronic' mechanism for the production of relativistic electrons than the `re-acceleration' mechanism, considering both the lack of steepening in the integrated spectrum and the spatial extent beyond the cold fronts; however the lack of spectral steepening may also be explained by continuous injection of relativistic electrons by undetected AGN jets.}
\item{The presence of the large-scale halo at high frequency hints at a spectral flattening with distance outside the cold fronts predicted by the `hadronic' mechanism; more data at intermediate frequencies with short baselines is required to confirm this along with more simulations to test if a similar effect could be caused by multiple AGN outbursts and/or cluster merger events.}
\item{The spectral steepening along the `tail' of the minihalo detected by other studies is more consistent with the `re-acceleration' mechanism; both mechanisms could therefore be important for producing relativistic electrons in different regions of the cluster.}
\end{enumerate}

\section*{Acknowledgements}

We thank the referee, Dr Rick Perley, for careful consideration of the manuscript and comments that helped us improve its presentation.  We thank the staff of the Mullard Radio Astronomy Observatory for their invaluable assistance in the commissioning and operation of AMI, which is supported by Cambridge University and ERC grant ERC-2012-StG-307215 LODESTONE.  This work is based on observations obtained with \emph{Planck} (http://www.esa.int/Planck), an ESA science mission with instruments and contributions directly funded by ESA Member States, NASA, and Canada.  The scientific results reported in this article are based in part on observations made by the \emph{Chandra} X-ray Observatory and published previously in cited articles.  This work utilized the R\={a}poi HPC facility at VUW.  The authors would like to thank Greg Willatt and David Titterington from Cavendish Astrophysics for computing assistance. PJE, KJ, TZJ acknowledge Science and Technology Facilities Council studentships. YCP acknowledges support from a Rutherford Discovery Fellowship.


\section*{Data Availability}

Calibrated AMI data used in this paper is available upon reasonable request from the authors.  \emph{Planck} data is publicly available through the \emph{Planck} Legacy archive (\url{https://pla.esac.esa.int/}) and the \emph{Chandra} pressure profile is publicly available through the ACCEPT database (\url{https://web.pa.msu.edu/astro/MC2/accept/}).  The VLA data used is publicly available through the VLA data archive (\url{https://archive.nrao.edu/archive/advquery.jsp}).



\bsp	
\label{lastpage}
\end{document}